\documentclass[reprint,aps,showpacs,pra,nofootinbib,superscriptaddress]{revtex4-2}
\usepackage{color}
\usepackage[dvipsnames]{xcolor}
\usepackage{newlfont}
\usepackage{graphicx}
\usepackage{amssymb}
\usepackage{amsmath}
\usepackage[caption=false]{subfig}
\usepackage{bm}
\usepackage{verbatim}
\usepackage[latin1]{inputenc}
\usepackage{bbm}
\usepackage{multirow}
\usepackage[thinlines]{easytable}
\usepackage{braket}
\usepackage{amsthm}
\usepackage{bbold}
\usepackage{subfig}
\usepackage{mathrsfs}
\usepackage[varg]{txfonts} 
\usepackage{amsfonts}

\newcommand{\ba}{\begin{array}}
	\newcommand{\ea}{\end{array}}

\newcommand{\mco}{\mathcal{O}}
\newcommand{\mck}{\mathcal{K}}
\newcommand{\txi}{\tilde{\xi}}
\newcommand{\tz}{\tilde{z}}
\newcommand{\tPsi}{\tilde{\Psi}}
\newcommand{\mcc}{\mathcal{C}}

\newcommand{\la}{\label}
\newcommand{\mbo}{\mathbf{O}}

\usepackage[pagebackref=false, colorlinks=true]{hyperref}
\definecolor{redish}{rgb}{0.7,0.2,0.0}  
\definecolor{bluish}{rgb}{0.2,0.5,0.8}
\hypersetup{linkcolor=red,          
	citecolor=teal,        
	filecolor=magenta,      
	urlcolor=blue}          

\begin{document}
	\title{Statistics and Complexity of Wavefunction Spreading in Quantum Dynamical Systems}

     \author{Yichao Fu}\email{yichao.fu@gm.gist.ac.kr}
 \affiliation{
	Department of Physics and Photon Science, Gwangju Institute of Science and Technology, 123 Cheomdan-gwagiro, Gwangju 61005, Republic of Korea}
 \author{Keun-Young Kim}\email{fortoe@gist.ac.kr}
  \affiliation{
	Department of Physics and Photon Science, Gwangju Institute of Science and Technology, 123 Cheomdan-gwagiro, Gwangju 61005, Republic of Korea}
 \affiliation{Research Center for Photon Science Technology, Gwangju Institute of Science and Technology, 123 Cheomdan-gwagiro, Gwangju 61005, Korea}
	\author{Kunal Pal}  \email{kunal.pal@apctp.org} 
\affiliation{
	Asia Pacific Center for Theoretical Physics, Pohang 37673, Republic of Korea}
\author{Kuntal Pal}\email{kuntalpal@gist.ac.kr}
\affiliation{
	Department of Physics and Photon Science, Gwangju Institute of Science and Technology, 123 Cheomdan-gwagiro, Gwangju 61005, Republic of Korea}

\begin{abstract}
We consider the statistics of the results of a measurement of the spreading operator in the Krylov basis generated by the Hamiltonian of a quantum system starting from a specified initial pure state. We first obtain the probability distribution of the results of measurements of this spreading operator at a certain instant of time, and compute the characteristic function of this distribution.  We show that the moments of this characteristic function
are related to the so-called generalised spread complexities,  and obtain expressions for them in several cases when the Hamiltonian is an element of a Lie algebra. Furthermore, by considering a continuum limit of the Krylov basis,  we show that the generalised spread complexities of higher orders have a peak in the time evolution for a random matrix Hamiltonian belonging to the Gaussian unitary ensemble. 
We also obtain an upper bound on the change in generalised spread complexity at an arbitrary time in terms of the operator norm of the Hamiltonian and discuss the significance of these results. 
\end{abstract}
		
\maketitle
	
\section{Introduction}
According to one of the fundamental principles of quantum theory, the description of possible results of measurements of a quantum mechanical observable,  mathematically described by a self-adjoint operator $\mbo$,  can be encoded in the statistics of the probability distribution function (PDF from now on) associated with that observable. The information contained in this probability distribution is equivalent to that encoded in the characteristic function of the distribution, where these two quantities are related by a Fourier transformation \cite{Barnett}. One can obtain moments of different order (and hence cumulants from them) from this characteristic function by taking derivatives with respect to a suitable parameter (which is essentially conjugate to the results of the measurement) \cite{Hofer}. These moments, for specific operators, contain information about the evolution generated by them.   To give one example, the first moment of the PDF is the expectation value of the operator with respect to the time-evolved state, while the second moment can be used to obtain the variance of the operator, and both these quantities (the expectation values and the variance), when the operator under consideration is the Hamiltonian, determine the so-called quantum speed limit  -  the maximum rate through which a quantum system can evolve \cite{Deffner:2017cxz}. 

In light of the above discussion, in this work, our goal is to study the statistics of a particular quantum mechanical operator - the spreading operator that measures the spread of an initial quantum state under unitary Hamiltonian evolution measured in a preferred basis, known as the Krylov basis \cite{viswanath1994recursion, Balasubramanian:2022tpr}. The basic notion of the Krylov basis and its recent extensive use to study quantum chaos and quantum information spreading have roots in the works of Parker et al. \cite{Parker:2018yvk}, where it was conjectured that for a quantum many-body system in the thermodynamic limit,  the so-called Lanczos coefficients ($b_{n}$) associated with a local operator show a maximal linear growth when the system is chaotic, with a logarithmic correction for one-dimensional systems. In this description,  starting from an initial operator and the  Hamiltonian, the Lanczos coefficients are the essential ingredients needed to construct an orthonormal basis in the operator Hilbert space when one chooses a suitable inner product. An important result related to this, but in the context of time evolution of a state rather than an operator, was proven in \cite{Balasubramanian:2022tpr}.  It was shown that for unitary Hamiltonian evolution of an initial quantum state,  when one defines a suitable cost to measure the spreading of the time-evolved state in a complete orthonormal basis, this cost is minimised up to a finite time when computed with respect to the Krylov basis generated by the Hamiltonian starting from the given initial state.  In this sense, as far as the description of the spread of an initial state under time-evolution is concerned,  this Krylov basis is somewhat `special'  compared to any other orthonormal basis which has the initial state as its first element, and therefore,  the cost becomes the `complexity' of the time-evolved state - the resulting quantity is consequently known as the spread complexity (SC). 

This preferred Krylov basis is constructed from the initial state and the system Hamiltonian by using the Lanczos algorithm, which gives two sets of coefficients $a_{n}$ and $b_{n}$, apart from the normalised and orthogonal Krylov basis,  as the output. The SC of the initial state can then be written as the expectation value with respect to the time-evolved state of the spreading number operator, which is diagonal in the Krylov basis.  Both the SC and the corresponding Krylov operator complexity have been used extensively in the recent past in various contexts, see  \cite{Barbon:2019wsy, Dymarsky:2019elm, Camargo:2022rnt,  Avdoshkin:2022xuw, Dymarsky:2021bjq, Rabinovici:2021qqt, Bhattacharjee:2022vlt, Erdmenger:2023wjg, Gautam:2023bcm, Du:2022ocp, Huh:2023jxt, Iizuka:2023fba,Bhattacharjee:2023dik, Chattopadhyay:2023fob, Kundu:2023hbk, Zhang:2023wtr, Nandy:2023brt, Gautam:2023pny,Bhattacharyya:2023dhp, Vasli:2023syq, Bhattacharyya:2023grv, spread1, spread2, Bhattacharya:2023yec, Gill:2023umm,  Loc:2024oen, Malvimat:2024vhr, Aguilar-Gutierrez:2024nau, Zhou:2024rtg, Cindrak:2024exj,  Nizami:2024ltk, Bhattacharya:2024hto, He:2024xjp, Camargo:2024deu, Baggioli:2024wbz, Takahashi:2024hex, Bhattacharya:2024szw,  Gill:2024acg, Suchsland:2023cmb, Li:2024ljz, Seetharaman:2024ket, Chowdhury:2024qaj, Balasubramanian:2024ghv,Ganguli:2024myj}. 
For a concise review of the basics of dynamics of time evolution in the Krylov basis,  and various applications, see the recent article \cite{Nandy:2024htc}. 

As we have mentioned previously, in this paper, we will explore the statistics of the spreading operator for a generic isolated quantum system evolved by a  time-independent Hamiltonian. Our motivation here is mainly two-fold. Primarily, we are interested in understanding how the information of the spreading of an initial state under Hamiltonian evolution in the Krylov basis is encoded in the statistics of the spreading operator. As we will see in subsequent sections, the moments of the characteristic function computed from the PDF of a measurement of the spreading operator naturally give rise to higher-order generalisations of the SC, called the generalised spread complexity (GSC). Then it is natural to expect that the statistics should contain more fine-grained information about the spreading than that SC can provide, e.g., the SC only tells us about the average position of a particle in a one-dimensional Krylov chain, whereas these higher moments have information about the variance of the position of the particle in that chain and so on \cite{Balasubramanian:2022tpr}. In fact, most of this paper is devoted to studying various properties of these higher-order moments and their interpretations. 

On the other hand, from a broader perspective, one of the most important questions that we want to ask  (and pursue further to its answer in this and future studies) is:  how one can quantify the  `complexity' of a process that has been performed on an isolated quantum system, and its relation to other information-theoretic and thermodynamic quantities, such as the work done on the system or the change of entropy? In fact, one can not always define these quantities straightforwardly and the definitions that have been used so far in the literature very much depend on the context.  E.g.,  it is well known from the exploration of quantum thermodynamics that there is no single observable that can be associated with the work done in a quantum process when quantified by the difference of energy before and after the realisation of the process in the so-called two-point measurement scheme \cite{Kurchan:2000rzb, Tasaki2000jarzynski, Talkner1}. The two-point measurement scheme,   one of the most popular protocols used to measure the work done on a quantum system, utilises two projective energy measurements on the system before and after a certain process, say a quantum quench.  Apart from the work done on the system,  as far as the complexity of the process is concerned,  we believe the SC is a good candidate in this regard  (e.g., see \cite{Gill:2023umm} for the definition of SC in a two-point measurement scheme).  Though this line of analysis will not be our primary focus in this paper, the analysis of statistics of the spreading operator for a single measurement, which we perform in this paper, is a reasonable starting point.   

\textbf{Outline:} The present paper is organised as follows. We first define the statistics of the measurement of a generic quantum mechanical observable in the next section and discuss the physical interpretations of the characteristic function of the PDF and its moments. In section \ref{statSO}, we then apply these definitions to a specific operator, namely the spreading number operator in the Krylov basis, and show that the moments of the characteristic function of the PDF of this operator are directly related to the GSC. Therefore, knowing the GSC, one can construct the characteristic function from its Taylor series expansion and hence the PDF itself by using the inverse Fourier transformation.  In the rest of the paper, we study various properties of these quantities. Specifically, in section \ref{propertiesGSC}, we discuss some general properties of these quantities, such as their behaviour in the energy basis, and show that the variance of the spreading operator can be interpreted as the magnitude of the velocity of the evolution generated by the Heisenberg time-evolved version of this operator.  In section \ref{contiumlimit}, we consider a continuum limit in the Krylov basis, which allows us to analytically show that GSCs of higher orders have peaks in their time-evolution before they reach the saturation at late times for Hamiltonians belonging to the Gaussian unitary random matrix ensemble and a generic initial state.  Section \ref{su2} deals with another analytical approach,  where, by assuming that the Hamiltonian is an element of a certain Lie algebra, we obtain exact analytical expressions for GSC, characteristic function, as well as the PDF of the wavefunction spreading. 
In section \ref{unitary}, we discuss the effect of a unitary transformation on the GSC and the PDF of the statistics of the spreading operator and use the unitary invariance of SC to provide a geometry interpretation of it, following \cite{Caputa1}.  Next, we derive a bound on the change of the GSC for a finite time in terms of the operator norm of the Hamiltonian to understand the importance of the complexity in quantifying the cost of the evolution 
in section \ref{bounds}.  Finally, we conclude in section \ref{summary} by providing a summary of the results presented in the manuscript and pointing out a few important future directions along which the results presented here can be extended. 

\section{Statistics of measurement of a quantum mechanical observable}\label{statO}
In this section, we first introduce the statistics of results obtained from a measurement of a general quantum mechanical observable at a certain time under a Hamiltonian evolution, and then obtain an expression for the characteristic function and its various moments.  We consider an isolated quantum mechanical system described by the Hamiltonian $H(\lambda)$, with $\lambda$ collectively denoting the parameters of the Hamiltonian, and a quantum mechanical observable $\mbo$ whose eigenstates and eigenvalues are denoted respectively with $\ket{O_n}$ and $O_n$. The initial state of the system at an initial time ($t=0$),  is taken to be the pure state $\ket{\psi_0}$, which is then evolved with the Hamiltonian $H(\lambda)$.  In this paper, we assume that the initial state is not an eigenstate of the Hamiltonian since, in that case, the evolution is trivial.  
 
At an interval of time $t$  after the start of the evolution, we perform a measurement of the observable and record the result of the measurement. If we  perform this operation a large number of times, the statistics of the eigenvalues of the operator $\mbo$ is given by  \cite{Hofer}
	\begin{equation}\la {ab}
		P(O, t)  = \sum_n | \bra{ O_n } e^{-it H} \ket{\psi_0}| ^2 \delta (O - O_n)~. 
	\end{equation}
From this expression for the distribution $P(O, t) $, and the definition of the characteristic function,
	\begin{equation}
		\chi_{\mbo} (u,t) = \int e^{-iu O}  P(O, t) ~ dO~,
	\end{equation}
we obtain the following expression for $\chi_{\mco} (u,t)$, 
	\begin{equation}\label{CFO2}
		\chi_{\mbo} (u,t) =\bra{\psi_0} e^{-iu \mbo(t)} \ket{\psi_0}~,~~~\text{with}~~\mbo(t)= e^{i H t} \mbo e^{-iHt}~.
	\end{equation}
Here $\mbo(t)$ is the Heisenberg time-evolved operator form of the initial operator $\mbo$, and $u$ is a parameter which is conjugate to the results of measurement $O$. From the expression in eq. \eqref{CFO2}, one can see that the characteristic function associated with the distribution of the measurement outcome of the observable $\mbo$ corresponds to a correlation function between the initial state $\ket{\psi_0}$ and the evolved version of it, where the evolution is with respect to the parameter $u$, and the evolution is generated by the time-evolved Heisenberg operator $\mbo(t)$.
Alternatively, it is possible to interpret this characteristic function as a correlation function between a time-evolved state $\ket{\Psi(t)}= e^{-it H} \ket{\psi_0}$ and 
its $u$-evolved version, $\ket{\Psi(t,u)}= e^{-iu \mbo} \ket{\Psi(t)}$, generated by the time-independent operator $\mbo$.  These two interpretations are entirely
equivalent to each other, and below we refer to both of them interchangeably according to the convenience of discussion. 

Before moving on, we note that a setup similar to what we have discussed above can also be used to describe a quantum system that is perturbed from its initial equilibrium through a so-called sudden quantum quench. In such a protocol, the parameters of an initial Hamiltonian $H_0(\lambda_0)$, describing an isolated quantum system,   are suddenly changed to a new set of values, so that the new  Hamiltonian is given by $H(\lambda)$. The initial state before the quench at $t=0$,  i.e.,  $\ket{\psi_0}$ is constructed from the eigenstate of the pre-quench Hamiltonian, and therefore, not an eigenstate of the post-quench Hamiltonian. The measurement of the observable $\mbo$ is performed at an interval of time $t$  after the quench \cite{Paraan}.  Furthermore, the interpretation of  $	\chi_{\mbo} (u,t) $ given above  is quite  similar to that of the distribution  of the work done  in a two-point 
measurement protocol, which is also a correlation function between the initial state and an evolved state generated by successive forward and backward evolution by the pre-quench and Heisenberg evolved post-quench Hamiltonians \cite{Talkner1, AS}.   
	
An observation similar to the one we have made above from eq. \eqref{CFO2} 
was used in  \cite{Gill:2023umm} to define the complexity of the spread of the initial state with respect to the parameter $u$, sometimes called the second time of evolution in the literature \cite{AS, Chenu:2017qdv}. Here, it is important to understand the roles played by the two parameters $t$ and $u$.   In the correlation function 
\eqref{CFO2}, the value of the parameter $t$, i.e., the evolution  generated by the Hamiltonian  $H$ determines the time-evolved operator 
$\mbo(t)$.  Therefore, the nature of this operator, and in a certain sense,  its `complexity'  has effects on the evolution of the corresponding characteristic function.  
	
Next, we consider the moments of the characteristic function (with respect to $u$). These are given by
\begin{equation}\label{avO}
		\bar{M}_n =  \frac{d ^n \chi_{\mbo} (u,t)}{du^n}\bigg|_{u=0}~= (-i)^n \int O^n  	P(O, t)  dO~=(-i)^n \braket{O^n}~. 
\end{equation}
Here we have used a bar over the notations for the moments to indicate that these are different from the moments of the Hamiltonian evolution, i.e., those obtained from the derivatives (with respect to $t$) of the characteristic function associated with the distribution of the work done \cite{spread2}. 
Using the expression for the distribution of the  eigenvalues of the observable $\mbo$ we can rewrite these moments as 
	\begin{equation}\label{moments1}
		\bar{M}_n   = (-i)^n  \sum_m | \bra{ O_m } e^{-it H} \ket{\psi_0}| ^2  O_m^n~= (-i)^n  \bra{\psi_0} \mbo^n (t) \ket{\psi_0}~ .
	\end{equation}
The Lanczos coefficients associated with the $u$-evolution can be computed using the standard relation between them and the moments of the auto-correlation function, i.e., $\bar{M}_n^*$ \cite{Gill:2023umm}; however, these will not be important for the purposes of the present paper. 
	
With this discussion of the statistics associated with the measurement of a general observable $\mbo$, we now consider a particular operator that is of interest to our present purpose, namely,  the spreading operator in the Krylov basis.
The goal of the rest of the paper is to find out the statistics of this operator in the setup described at the beginning of this section and study various properties associated with these statistics.
	
\section{Statistics of spreading operator in  the Krylov basis} \label{statSO}
In this section, we first define the statistics associated with the so-called spreading number operator,  which we refer to here as the spreading operator from now on for convenience.   
This operator is used to define a measure of the spreading of an initial state ($\ket{\psi_0}$) with time evolution
generated by the Hamiltonian $H$ in subspace generated by $H$ starting from $\ket{\psi_0}$, known as the Krylov subspace.\footnote{In this paper, the notation $H$ always denotes the Hamiltonian that generates the time evolution for $t>0$  (or the post-quench Hamiltonian when the setup describes a sudden quench protocol, though we do not consider quenches explicitly in this paper). In the rest of the paper, we shall omit the prefix ``post" for convenience whenever no confusion arises, i.e., when the pre-quench Hamiltonian (always denoted by $H_0$) does not play any active role. }  The  spreading operator is defined as 
\begin{equation}\label{spoper}
		\mck = \sum_n n \ket{K_n} \bra{K_n}~,
\end{equation}
where $\ket{K_n}$ denotes the Krylov basis constructed from the initial state $\ket{\psi_0}$ and the Hamiltonian $H$, with $n=0,1,2.\cdots$.  Here $n$ is the eigenvalue of the operator $\mck$, with $\ket{K_n}$ being the corresponding eigenvectors. The summation over $n$ runs over the dimension of the Krylov basis, which we denote as  $L$.
Note that $\mck$ is the operator which is diagonal in the Krylov basis, and each element of this basis is related to the initial state through the relation $\ket{K_n}=
\mathcal{P}_n(H) \ket{\psi_0}$, where $\mathcal{P}_n(H) $ denote normalised and orthogonal polynomials of degree $n$ \cite{Dehesa,Muck:2022xfc}. 

In this approach of studying time evolution in the Krylov basis, it is possible to map any complicated unitary
quantum dynamics to a hopping problem in a one-dimensional 
chain of lattice. Then, the eigenvalues of the spreading 
operator (the spreading number $n$) can be thought of as the position of a fictitious particle on this one-dimensional Krylov chain, and therefore, a measurement of this operator 
gives the position of the particle on the chain at an instant of time.

For the time-evolved state  $\ket{\Psi(t)}=e^{-it H}\ket{\psi_0}$, the SC is given by the expectation value of the spreading operator in this time-evolved state,
	\begin{equation}
		\mathcal{C}(t)=  \bra{\Psi(t)} \mck \ket{\Psi(t)}~=  \bra{\psi_0} \mck (t) \ket{\psi_0}.
	\end{equation}
A measurement of the operator $\mck$ will produce a value $n$.  Since the initial state is the first state of the Krylov basis ($\ket{\psi_0}=\ket{K_0}$), a measurement of the SC operator before the start of the evolution will produce a value of zero, i.e., the spreading of the initial state in the Krylov basis is zero. 
Therefore, the statistics associated with a post-evolution projective measurement outcome of $j$  of  the operator $\mck$  can be written as 
	\begin{equation}\label{SOsta}
		P(j, t)  = \sum_n | \bra{ K_n} e^{-it H} \ket{\psi_0}| ^2 \delta (j-n)~=  \sum_n   |\phi_n(t)|^2 ~ \delta (j-n)~,
	\end{equation}
where  $ \phi_n(t)= \braket{K_n|\Psi(t)} $ are the expansion coefficients of the time-evolved state in the Krylov basis.  This quantity essentially gives the distribution of the spreading number of the time-evolved state in the Krylov basis at an instant of time $t$ when the measurement of the operator $\mck$ is performed on the system. 
	
We first consider the characteristic function associated with this distribution, which, according to the general formula in eq. \eqref{CFO2},  is  given by
	\begin{equation}\label{CFSC}
		\chi_{\mck} (u,t) =\bra{\psi_0} e^{-iu \mck(t)} \ket{\psi_0}~,~~~\text{with}~~\mck(t)= e^{i H t} \mck e^{-iHt}~.
	\end{equation}
	Here $u$ is the variable conjugate to $j$ in the corresponding Fourier transformation.  This expression for the characteristic function can be rewritten as
	\begin{align}\label{CFSC2}
		\chi_{\mck} (u,t)  = \sum_{m=0}^{\infty} \frac{(-iu)^m}{m!} ~ \bra{\psi_0} \Big(\mck(t)\Big)^m \ket{\psi_0} \\
		= \sum_{m, n}  \frac{(-iu)^m}{m!} n^m  |\phi_n(t)|^2~.
	\end{align}
	Therefore,  we see that each term in the summation above contains  factors of the form 
	\begin{equation}
		\bra{\psi_0} \Big(\mck(t)\Big)^m  \ket{\psi_0}= \sum_n n^m |\phi_n(t)|^2~. 
	\end{equation}
To understand the significance of these quantities,  consider  the expansion of a time-evolved state with respect to  an orthonormal and complete basis  $\ket{B_n}$, $n=0,1,\cdots$, and define the following
quantity, 
	\begin{equation}\label{cost}
		\mathcal{C}_{\mathcal{B}} (t) = \sum_n c_n | \braket{\Psi(t) | B_n}|^2~= \sum_n c_n p_B (n,t)~,
	\end{equation}
which is usually referred to as the cost function and is a measure of the spreading of the time-evolved state in the orthonormal basis under consideration \cite{Balasubramanian:2022tpr}. Here $c_n$ denotes an increasing sequence of positive real numbers.  As shown in \cite{Balasubramanian:2022tpr}, such a cost function is minimised over
a finite interval of time around $t=0$ when computed in terms of the Krylov basis, i.e., when $\ket{B_n}=\ket{K_n}$.   When the real numbers $c_n$ are taken as $c_n=n^{m}$, with $n=0,1, .\cdots$, and $m=1,2,3 \cdots$, the above expansion in the Krylov basis reduces to
	\begin{equation}\label{geSC}
		\mathcal{C}_m (t) = \sum_n n^m | \braket{\Psi(t) | K_n}|^2~= \sum_n n^m|\phi_n(t)|^2 ~,
	\end{equation}
and all these quantities are minimised during a finite time interval after the initial time compared with any complete orthonormal basis.  
We notice that these quantities are similar to the ones studied in \cite{Fan:2023ohh} in the context of operator growth,   and subsequently dubbed as the generalised Krylov complexity. Therefore, here we call $\mathcal{C}_m (t)$ the generalised spread complexity  (GSC, or GSCs when referring to all of them, i.e., for all $m$, collectively). When $m=1$ we get back the expression for SC.
Now the expression for the characteristic function associated with the distribution of spreading from eq. \eqref{CFSC2} can be rewritten as
	\begin{align}\label{CF2}
		\chi_{\mck} (u,t)  
		= \sum_{m}  \frac{(-iu)^m}{m!}  \mathcal{C}_m(t)~.
	\end{align}
Using this relation along with the definition of the moments of the characteristic function in eq. \eqref{moments1}, we see that the moments of the characteristic function associated with the distribution of the eigenvalues of the spreading operator are related to the GSC:
	\begin{equation}\label{GSC=Mm}
		\mathcal{C}_m (t) = \sum_n n^m|\phi_n(t)|^2 ~=\frac{1}{(-i)^m}\bar{M}_m (t) ~.
	\end{equation}
In other words, knowing the GSC, one can obtain the characteristic function from its Taylor series expansion with respect to the parameter $u$, and hence the PDF  in eq. \eqref{SOsta} itself by taking an inverse Fourier transformation. 
 
One way of obtaining all the GSCs from the expression for the expansion coefficients  $ \phi_n(t)$  is the following. Define the function
	\begin{equation}\label{Generating}
		G(\eta,t)= \sum_{n} e^{\eta n}  |\phi_n(t)|^2 ~,
	\end{equation}
where $\eta$ is an auxiliary variable.  Putting $\eta=-iu$ in this definition,  and comparing it with eq. \eqref{CF2}, we see that this is related to the characteristic function through the relation $G(-iu,t)=\chi_{\mck} (u,t)  $.\footnote{Notice that at $\eta=0$, $G(\eta=0,t)=1$ due to probability conservation. }
Considering a Taylor series expansion of $G(\eta, t)$ with respect to $\eta$ around $\eta=0$, it can be seen that, $m$-th order derivative of this function 
at $\eta=0$ is equal to the $m$-th order GSC, 
\begin{equation} \label{generating}
		\frac{d^m G(\eta,t)}{d \eta^m}\Bigg|_{\eta=0} = \sum_n n^m|\phi_n(t)|^2 = 	\mathcal{C}_m (t) ~.
\end{equation}
As we shall see, in many examples, it is possible to directly compute the function $G(\eta,t)$ from the knowledge of the expansion coefficients, so that the relation in 
eq. \eqref{generating} provides a convenient way of obtaining  GSC for any value of $m$. 
	
To understand the significance of these GSCs further, we note from eq. \eqref{avO} and  eq. \eqref{geSC} above that the GSCs are actually averages of different powers of the spreading  \footnote{Here spreading is with respect to the Krylov basis, measured by the spreading number $n$. Unless otherwise stated, in this paper, we always use the Krylov basis to quantify the spreading of a time-evolved state.} in a measurement protocol where the operator measured is the spreading operator in the Krylov basis generated by the Hamiltonian $H$, i.e., 
	\begin{equation}
		\braket{n^m} = \mathcal{C}_m (t) =  \sum_n n^m|\phi_n(t)|^2 ~. 
	\end{equation}
These can be written as the expectation value of the generalised spreading operator, 
	\begin{equation}
		\mck_m = \sum_n  n^m \ket{K_n} \bra{K_n}~
	\end{equation}
	with respect to the time-evolved state, $\braket{n^m} = \mathcal{C}_m (t)  = \braket{\Psi(t)| \mck_m | \Psi(t)}$. 
	In particular, the average spreading number is given by the SC itself, while the variance of the spreading is 
	\begin{equation}\label{variance}
		(\Delta n(t))^2 = \braket{n^2} - \braket{n}^2~=  \sum_n n^2|\phi_n(t)|^2 - \Big(  \sum_n n|\phi_n(t)|^2\Big)^2~.
	\end{equation}
	
Proceeding in a manner similar to what we have described so far in this section,  one can also obtain the statistics of the generalised spreading operator, $\mck_m$. 
The characteristic function of this distribution can be written as (from eq. \eqref{CF2} with $\mck_m(t)= e^{i H t} \mck_m e^{-iHt}$)
	\begin{equation}\label{CFSCm}
		\begin{split}
			\chi_{\mck_{m}} (u,t) =\bra{\psi_0} e^{-iu \mck_m(t)} \ket{\psi_0}~\\
			= \sum_{j, n}  \frac{(-iu)^j}{j!} n^{mj}  |\phi_n(t)|^2~.
		\end{split}
	\end{equation}
Clearly, depending upon the value of $m$, only averages of certain powers of the spreading operator contribute to the above expression. E.g., when $m=2$, only averages of the even powers of the spreading operator, i.e., $\braket{n^{2j}}$ determine the statistics of the operator $\mck_2$. 

\section{General Properties of spreading operator statistics}\label{propertiesGSC}
It should be clear from the discussions presented in the last two sections that the statistics of the fluctuation of the spreading operator can be obtained from the knowledge of the GSCs of the time-evolved state generated by the Hamiltonian $H$. Therefore, in the rest of this paper, we shall study various properties of GSCs. 
Specifically, in this section, we first write down the expressions for the PDF in eq. \eqref{SOsta} and its moments (and hence, the GSCs)  in terms of the energy basis, which will be later used in section \ref{contiumlimit} to derive expressions for the GSCs in the continuum limit.  Furthermore, in section \ref{geometric}, we shall discuss a geometric interpretation of this variance as the magnitude of the velocity of the evolution along the unitary curve generated by the spreading operator, and in section \ref{bounds}, we shall derive an expression for the upper bound of the total change of these GSCs during a finite time under a Hamiltonian evolution. 

The properties of GSCs we discuss in this section are quite general and valid for any arbitrary initial pure state and system Hamiltonian. In the latter part of this paper, we shall consider some specific models where one can compute these quantities analytically and therefore, explicitly verify different properties discussed here. 

\subsection{Probability distribution and Moments in the energy basis}\label{energybasis} 
We begin by writing down the expression for the probability distribution and its moments (or the GSCs)  in terms of the energy eigenbasis of the Hamiltonian, since these will be useful in studying the long-time average and continuum limit of these quantities. We denote the eigenstates and eigenvalues of  $H(\lambda)$ as $\ket{E_a (\lambda)}$ and $E_a(\lambda)$, respectively.  For simplicity, we assume that the energy spectrum is non-degenerate. 
	
First, consider the expression for the probability distribution in \eqref{SOsta}. This can be rewritten in terms of the energy basis as 
	\begin{equation}
		P(j,t)  =   \sum_{n,a,b}  ~~ \rho_0 (E_a, E_b)~  \mathbf{P}_n(E_a, E_b) ~e^{i(E_a-E_b)t} ~ \delta(j-n)~,
	\end{equation}
	where $ \rho_0 (E_a, E_b) = \braket{E_b | \psi_0} \braket{\psi_0|E_a} $ and $\mathbf{P}_n(E_a, E_b)= \braket{E_b | K_n}\braket{K_n|E_a}  
	= \braket{E_b | \mathbf{P}_n |E_a}  $ denote, respectively,  the elements of the initial state density matrix and the projection operator on the $n$-th Krylov basis in the energy basis. Thus, $\mathbf{P}_0(E_a, E_b)= \rho_0 (E_a, E_b)$. 
When written  in terms of the orthogonal polynomials mentioned earlier, we have,
	\begin{equation}
		P(j,t)  =   \sum_{a,b}  ~~  P(j,E_a, E_b) ~\rho_0 (E_a)  \rho_0 (E_b)~e^{i(E_a-E_b)t} ~ ,
	\end{equation}
	where we have defined, 
	\begin{equation}
		P(j,E_a, E_b) = \sum_n  \delta(j-n)  \mathcal{P}_n(E_a) \mathcal{P}_n(E_b)~, 
	\end{equation}
	$\rho_0 (E_a)= |\braket{\psi_0| E_a}|^2$, and $\mathcal{P}_n(E_a)=\braket{E_a | K_n}/ \braket{E_a | K_0}$ (so that $\mathcal{P}_0(E_a)=1$). 
One can follow similar steps to write down the expressions for the GSC in eq. \eqref{geSC}  as 
	\begin{equation}
		\mathcal{C}_m (t)  =   \sum_n n^m  \sum_{a,b} \braket{E_a | K_n} \braket{K_n | E_b} ~ \rho_0 (E_a, E_b)~e^{i(E_a-E_b)t} ~,
	\end{equation}
which can be further rewritten as 
	\begin{equation}\label{GSC2}
		\mathcal{C}_m (t)	 = \sum_{a,b}  A_m (E_a, E_b) \rho_0 (E_a, E_b)~e^{i(E_a-E_b)t}  ,
	\end{equation}
with $ A_m (E_a, E_b)= \bra{E_a} \mathcal{K}_m \ket{E_b}= \sum_n n^m  \mathbf{P}_n(E_a, E_b)$, indicating the elements of the generalised spreading operator in the energy basis. An alternative expression of GSC, which can be easily obtained from the above, is,
	\begin{equation}\label{GSC3}
		\mathcal{C}_m (t)=  \sum_{a,b} ~  J_m (E_a, E_b)   \rho_0 (E_a)  \rho_0 (E_b)~  e^{i(E_a-E_b)t} ~,
	\end{equation}
where 
	\begin{equation}
		J_m (E_a, E_b)  = \sum_n  n^m \mathcal{P}_n(E_a) \mathcal{P}_n(E_b)~. 
	\end{equation}
This form for the GSC is particularly suitable for studying the time dependence of this quantity in the continuum limit, and when the Hamiltonian $H$ has a continuous spectrum.  
We discuss the behaviour of the  GSC in this limit in more detail in section  \ref{contiumlimit}. 
	
As an aside, we note that, apart from the moments of the distribution of the spreading, another important quantity associated with this distribution is its entropy,  defined by 
	\begin{equation}\label{ent}
		\mathcal{S}_\mathcal{K}(t)=-\sum_j  P(j,t) \ln P(j,t)~.
	\end{equation}
The  Shannon entropy of the work distribution in a two-point measurement protocol has been studied in detail recently in  \cite{Kiely:2022ckd}, 
and its importance in understanding the localisation transitions has been discussed.  A detailed study of this quantity is beyond the scope of the present paper. However, it will be interesting to analyse different properties of the entropy in \eqref{ent} and compare it with the entropy of the uncollected distribution, as in \cite{Kiely:2022ckd}.    
 
\subsection{A geometric interpretation of the variance of spreading }\label{geometric}
Before concluding this section we briefly discuss a geometric interpretation of the variance of the spreading introduced above.  Consider the state $\ket{\psi(u)}=e^{-iu \mck(t)} \ket{\psi_0}$,
whose evolution in the parameter  $u$ is generated by $\mathcal{K}(t)$.  Now, the uncertainty of the Heisenberg evolved operator $\mck(t)$ in this $u$-evolved state, 
	\begin{equation}
		(\Delta \mathcal{K}(t))^2= \bra{\psi(u)}\mck(t)^2 \ket{\psi(u)}-  \bra{\psi(u)}\mck(t) \ket{\psi(u)}^2~,
	\end{equation}
can be seen to equal the variance of the spreading $(\Delta n)^2$. 

Following \cite{Anandan}, we can associate a notion of geometry with this evolution.  Consider the projective Hilbert space ($\mathcal{P}$), i.e., the set of rays associated with a Hilbert space $\mathcal{H}$.  The evolution of the state  $\ket{\psi(u)}$ projects a curve (which we denote as $\mathcal{U}(u))$ in the projective Hilbert space. It is well known that one way of introducing  a concept of distance between two close-by states $\ket{\psi(u)}$ and $\ket{\psi(u+du)}$ on this curve is through the Fubini-Study metric, such that the line element can be written as 
	\begin{equation}
		ds^2= N \Big[1- \big| \braket{\psi(u) |  \psi(u+du)}\big|^2\Big]~,
	\end{equation} 
where $N$ is a constant, which we subsequently take to be 1.\footnote{Different authors make different choices for this constant, e.g., in ref \cite{Anandan} it was taken to be $4$, whereas the authors of  \cite{Dodonov}  use $N=2$. } By expanding the state $\ket{\psi(u+du)}$  up to the second order
	in $du$, we can reach an analogous relation as in \cite{Anandan}, between the infinitesimal path length $ds$  and the uncertainty of the operator $\mck(t)$ 
	defined above,  $ds= \Delta \mck(t) ~du$. Therefore, the variance of the spreading is equal to the magnitude of the velocity of evolution along the unitary curve $\mathcal{U}(u)$ generated by the time-evolved spreading operator.  We also notice that, even though both $\braket{n}$ and $\braket{n^2}$ are individually minimised when computed with respect to the Krylov basis, the variance of the spreading may not be the minimum. Therefore, the speed of the evolution is not guaranteed to be the minimum in the Krylov basis. 

 \subsection{Loschmidt-echo of $u$-evolution}
 Another interesting quantity one can define from the characteristic function in eq. \eqref{CFSC} is the survival probability of the initial state after the $u$-evolution generated by the operator $\mck(t)$, i.e., the corresponding Loschmidt echo (which we call $u$-Loschmidt echo in the following to distinguish it from the Loschmidt echo of $t$-evolution). It is easy to show that this is related to the GSC through the relation  
	\begin{equation}
		\mathcal{T}(u,t) = |\chi_{\mck}(u,t)|^2  = \sum_{n,m} (-1)^n  \frac{(iu)^{n+m}}{n! m!}  \mathcal{C}_n(t)  \mathcal{C}_m(t)~.
	\end{equation}
For fixed values of $t$, when this quantity is zero, the $u$-evolved state becomes orthogonal 
to the initial state, while when  this returns to the initial value, the $u$-evolved state 
coincides with the initial state.  
For the analytical examples we consider later in this paper, it is possible to exactly evaluate the summation and obtain an expression for the  $u$-Loschmidt echo in terms of the SC (see eq. \eqref{LEsu2} in section \ref{su2} for an example of such a relation when the  Hamiltonian $H(\lambda)$ is an element of the $su(2)$ Lie algebra). 
As we shall see, for these examples, the GSC of order $m$ can be written as a polynomial function of the SC of degree $m$.  In such cases, the time instances when the SC vanishes, all higher order  $\mcc_m(t)$s also vanish, so that the $u$-Loschmidt echo also goes to zero at these instances. 

\section{Continuum limit of the generalised spread complexity} \label{contiumlimit}
In general, for a given initial state and a Hamiltonian,  it is often quite difficult to analytically solve the discrete Schrodinger equation for the associated  Lanczos coefficients to obtain the expansion coefficients $\phi_n(t)$, and hence the SC.  One way we can proceed analytically to obtain these is by considering the so-called continuum limit, i.e., by assuming that the discrete index $n$  can essentially be taken as a continuous coordinate, say $x$.  	Assuming that the Lanczos coefficients and the wavefunctions $\phi_n(t)$ depend smoothly on $x$ one can transform the discrete Schrodinger equation for these wavefunctions to a first-order differential equation, and subsequently solve it to obtain the wavefunctions. These wavefunctions can then be utilised to obtain an expression for the SC in the continuum limit.  The expressions for the wavefunctions and the SC obtained from computation using the continuum limit often turn out to be quite useful, see e.g.,  \cite{Barbon:2019wsy, Muck:2022xfc, Alishahiha:2022anw, Erdmenger:2023wjg}. 


 In this section, we discuss the behaviour of the GSC  in the continuum limit.
Here, our goal is to find out the time dependence of the GSC of different orders in this approximation from a given form for the initial density matrix and the pattern of the Lanczos coefficients. Specifically,  we show that for a random matrix Hamiltonian belonging to the Gaussian unitary ensemble (GUE), there exists a peak in the expression of the GSC of all orders.  The fact that a peak exists, even in the continuum limit,  is important since it has been used recently by several authors \cite{Balasubramanian:2022tpr, Erdmenger:2023wjg, Camargo:2024deu, Baggioli:2024wbz}  as an indication of the chaotic nature of the Hamiltonian generating the time evolution.  Furthermore, as we shall see, these peaks are more prominent in higher-order GSCs than in the first-order SC.  Therefore,  one can use GSCs as sharper indicators of the chaotic nature of the spectrum of the Hamiltonian generating the evolution. 
	
Assume that we can replace the discrete index $n$ by a continuous coordinate $x$ by introducing a cutoff $\epsilon$ (with $\epsilon \ll 1)$ so that $x_n= \epsilon n$, and the Lanczos coefficients ($b_n$, and $a_n$) as well as the orthogonal 
polynomials and the wavefunctions are smooth functions of this continuous coordinate $x$. We denote these functions as  $b(x_n)=b_n, a(x_n)=a_n$,  $\mathcal{P}_n(E)= i^{-n} \mathcal{P}(E,x_n)$ and $\phi_n(t)= i^{-n} \phi(t,x_n)$. With these assumptions, the expression for the GSC, up to an overall constant normalisation factor, is 
given by the one in eq. \eqref{GSC3}, while the expression for $J_m(E_a, E_b)$ in the continuum limit is given by the following integral \footnote{This is the generalisation for the expression for the case $m=1$ given in   \cite{Erdmenger:2023wjg}. Since, in the continuum limit, the polynomials lose their normalisation, one needs to normalise them properly, as we do below. See ref. \cite{Erdmenger:2023wjg} for further details.}
\begin{equation}\label{Jmcont}
		J_m(E_a, E_b) = J_m(\omega) =  \frac{2}{\epsilon^{m-1}} \int_{0}^{y(\epsilon L)} dy ~ x^m(y) ~b(y)~ \cos (\omega y)~,
\end{equation}
with $\omega= E_a-E_b$. The coordinate $y(x)$ can be obtained by integrating the relation $dy= dx / (2\epsilon b(x))$ for a given profile 
of the second set of Lanczos coefficients, $b(x)$. We first obtain the expression of $J_m(E_a, E_b) $
for a given shape of the Lanczos coefficients in the continuum limit.  
To this end, we use the following result, which states that  for a Hamiltonian belonging to the classical Gaussian ensembles, 
and an initial state which has the form $(1,0,0, \cdots,0)^T$ when written in terms of a basis with respect to which the Hamiltonian is a random matrix belonging to the ensemble under consideration,  the Lanczos coefficients are independently distributed random variables \cite{dumitriu2002matrix,Balasubramanian:2022dnj} (see also \cite{Trotter1984EigenvalueDO}). 

In the limit of large $L$, it is possible to obtain a simple expression for the mean 
of the distribution of the Lanczos coefficients, and one can then use this as a pattern for the Lanczos coefficients in the continuum limit as a first approximation, since variances are negligible compared to the averages \cite{Balasubramanian:2022dnj, Erdmenger:2023wjg}. Therefore, the expressions for the Lanczos coefficients we use in the continuum limit are obtained from the expressions for the respective means of distribution and are given by  
\begin{equation}\label{LCrmt}
		a(x)=0~,~~b(x)= \sqrt{1-\frac{x}{\epsilon L}}~.
\end{equation}
Before using these approximations for the Lanczos coefficients to compute  $J_m(E_a, E_b) $, we note the following related points. 1. For the above expressions to be valid, one has to scale the Hamiltonian in such a way that $b_1=1$.  For a general non-scaled Hamiltonian, this is not usually the case. 
2. The average of the distribution of $a_n$ is always zero, irrespective 
of the values of $L$ and $n$. 
3. The variances of the distributions of both $a_n$ and $b_n$
scale as  $1/L$ in the leading order for large $L$, and therefore, can be neglected compared to the means. 
4. As we have mentioned above, these expressions for the mean of the distribution of the Lanczos coefficients are valid for a reference state of the form $(1,0,0, \cdots,0)^T$  written in an appropriate basis (also states related to this one through appropriate transformations depending on the ensemble under consideration). However,  it was pointed out  in \cite{Erdmenger:2023wjg}  that one can use the expressions in \eqref{LCrmt}
 as an approximation for Lanczos coefficients associated with the maximally entangled state (defined in a double-copy Hilbert space) and a Hamiltonian belonging to the Gaussian $\beta$-ensemble, since \eqref{LCrmt} is a fixed point of the Lanczos algorithm for this particular initial state. We use this observation in the following. 

 From the expression for $b(x)$ in \eqref{LCrmt} we first obtain  $y(x)= L \Big(1-\sqrt{1-\frac{x}{\epsilon L}}\Big)$, i.e., $x(y)= \frac{\epsilon}{L} y (2L-y)$, and subsequently utilising it we derive the expressions for different orders of $J_m(\omega)$.  For example, when $m=2$, we have  (with $l=L \omega$)
	\begin{equation}
		J_2(\omega)  = \frac{2 \epsilon}{l^3 \omega^3} \Big(120-48 l^2-\big(120+12  l^2 +  l^4\big) \cos l \Big)~.
	\end{equation}
This  is an even function of $\omega$, and in the limit, $\omega \rightarrow 0$, it is regular and has an expansion of the form (with $\epsilon$ taken equal to $1/L$):
	$J_2 (\omega \rightarrow 0)= L^2 \Big(\frac{1}{3}-\frac{47}{840} l^2+ \frac{17}{7560} l^4+ \mathcal{O}(l^6)\Big)$.
Furthermore, it is easy to check that all the functions $J_m(\omega)$ are proportional to $\epsilon$, and they are localised around $\omega=0$, having a width which decreases with the increasing dimension of the Krylov subspace. 

In the continuum limit, the expression for the GSC takes the following form 
\begin{equation}
    \mathcal{C}_m (t) = \frac{1}{L^2}  \sum_{a,b} ~ J_m (E_a, E_b)~ e^{i(E_a-E_b)t} ~,
\end{equation}
where $J_m (E_a, E_b)$, in this limit is given by the expression in \eqref{Jmcont}. 
To properly normalise this expression for GSC, we divide it by the total probability, which can be written as  
	\begin{equation}
		\mathbb{P} =  \frac{1}{L^2}  \sum_{a,b} ~ P (E_a, E_b)~ e^{i(E_a-E_b)t} ~.
	\end{equation} 
Here $P (E_a, E_b)$ in the continuum limit is given by \cite{Erdmenger:2023wjg}
	\begin{equation}
		P (E_a, E_b)=P (\omega) = 2 \epsilon \int_{0}^{y(\epsilon L)} dy ~b(y)~ \cos (\omega y)~.
	\end{equation}
For the Lanczos coefficients given in eq. \eqref{LCrmt} we get the expression for this function to be $P(\omega)= \frac{2 \epsilon}{l \omega} (1-\cos l)$. 

Next, we compute the ensemble average of these quantities for the GUE,   for which, the one-point function of the spectral density (i.e., here $\braket{\rho(E)}$), in the limit of large $L$, is given by the semi-circle law, $\rho_{\text{sc}}(E)=\frac{L}{2\pi}\sqrt{4-E^2}$, while, the two-point correlation function is given by the sine kernel formula (with $E=(E_a+E_b)/2$) \cite{Mehta, Cotler:2016fpe}, 
	\begin{equation}
		\begin{split}
			\braket{\rho(E_a)\rho(E_b)} = \braket{\rho(E)} \delta(\omega) \\
			+ \braket{\rho(E_a)} \braket{\rho(E_b)} \Bigg[1
   -\frac{\sin^2 \Big(\pi \braket{\rho(E)} \omega\Big)}{\Big(\pi \braket{\rho(E)} \omega\Big)^2}\Bigg]~.
		\end{split}
	\end{equation}
For calculating the ensemble averages  $\braket{\mathcal{C}_m}$ and $\braket{\mathbb{P}}$ we use the so-called box approximation \cite{Cotler:2016fpe,Cotler:2017jue,Liu:2018hlr}, which is equivalent to the procedure where one replaces both $E_a, E_b$ by the average $E$ in the connected part, and also approximate, in the sine kernel, the one-point function $\braket{\rho(E)}= \rho_{\text{sc}}(E)$, by its value at $E=0$, i.e.,  $\rho_{\text{sc}}(0)=L/\pi$. Here, we describe the computation of $\braket{\mathcal{C}_2(t}$, and one can follow an exactly similar procedure to compute $\braket{\mathbb{P}}$, as well as  $\braket{\mathcal{C}_m(t}$ for $m>2$.   
	
The procedure we follow is quite similar to the one used in \cite{Iliesiu:2021ari} to compute the length of the Einstein-Rosen bridge in models of dilaton gravity.  To evaluate the integral over $\omega$ in $\braket{\mathcal{C}_2(t)}$, we first notice that the integrand is a regular function of $\omega$. In fact, each of the terms, $J_2(\omega)$,  and the two-point function with sine kernel are separately regular. \footnote{One can check using the expression in \eqref{Jmcont} that this statement is true for higher order $J_m(\omega)$ as well. } Therefore, in the complex $\omega$ plane, we can deform the contour slightly without changing the value of the integral. Here we shift it to the upper half-plane. Now, we expand the sine-kernel in terms of exponentials, and for each term, close the semicircle in the upper or lower half of the complex plane so that the contribution from the semicircle arc is zero when extended to the complex infinity. Since, we need to consider terms of the form $\frac{e^{i  \# \omega}}{\omega^n}$, we can ensure that the contribution from the semi-circle vanishes when the contour is closed in the lower half-plane for $\#<0$, while it is closed in the upper half-plane when $\#>0$. Each of the terms in the expanded integral now has a pole at $\omega=0$ of order either $2,4,6$ or $8$; however, only the terms for which the contour is closed in the lower half-plane, i.e., $\#<0$,  have non-zero contributions to the $\omega$ integral due to the pole.  Since $\#<0$ can occur in three different ways, one needs to consider four time intervals separately, respectively,  $t<L$, $L \leq t < 2L$, $2L \leq t< 3L $ and $t \geq 3L$. Finally, we can straightforwardly perform the integral over $E$ (which has a limit from $-2$ to $2$, since the one-point function is given by the semi-circle formula) for each of the integrals and obtain expressions for $\mathcal{C}_2(t)$ in these time intervals. 
	
One can follow an analogous procedure to obtain the expressions for $\mathbb{P}$ in these intervals as well. The final result for the normalised, ensemble-averaged  GSC of second order, which we denote as $\braket{\mathcal{C}^N_{2}(v)}= \braket{\mathcal{C}_{2}(v)}/(L^2 \braket{\mathbb{P}(v)})$, in terms of the variable  $v=t/L$,  is given by
	\onecolumngrid
	\begin{equation}
		\braket{\mathcal{C}^N_{2}(v)}= \left\{ \begin{array}{rcl}
			\frac{(19\pi/7)+640v^2-1280v^3+(800+10 \pi) v^4-(160+12 \pi) v^5 + 5\pi v^6 -(5\pi/7)v^7}{160+5 \pi -160 v + 15 \pi v^2 - 5 \pi v^3}~,	& v<1 \\ 
			\frac{-19-35v-70v^4+84v^5-35v^6+5v^7}{35(-1-3v-3v^2+v^3)}~, & 1 \leq v \leq 2\\
			\frac{(-6637/7)+2565 v - 2880 v^2+1760v^3-630v^4+132v^5-15v^6+(5/7)v^7}{-75+135v-45v^2+5v^3}~,&  2 \leq v \leq 3\\
			\frac{1}{3}~, & 3 \leq v~.
		\end{array}\right.
	\end{equation}
	
\twocolumngrid
	
	\begin{figure}[h!]
		\centering
	\includegraphics[width=3.3in, height=2.2in]{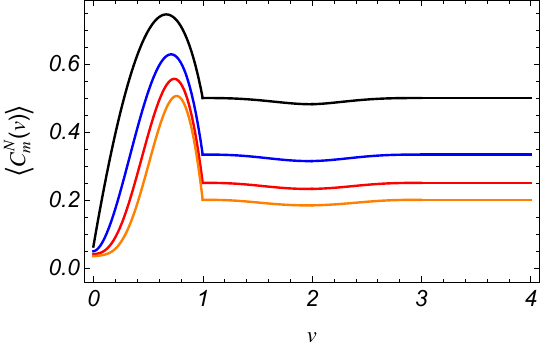}
  \caption{Plots of GSCs of order $m=2$ (blue), $m=3$ (red), $m=4$ (orange) in the continuum limit as a function of  $v=t/L$. For comparison, we have also plotted the SC in this limit (shown by the black curve).}
		\label{fig:Cmcont}
	\end{figure}
	
A plot of this function is shown in Fig.  \ref{fig:Cmcont}. In this plot, for comparison, we have also plotted the expression for the SC derived in \cite{Erdmenger:2023wjg} in the continuum limit, as well as the GSC of order $m=3,4$ obtained following the same procedure as above. The overall behaviour of all of these curves with time is quite similar, namely, all of them follow the general behaviour of rise, peak, and saturation. However, there are certain details where the plots differ from each other. Specifically, the saturation values, as expected, are different, e.g.,  the 2nd order normalised  GSC ($\braket{\mathcal{C}^N_{2}(v)}$) saturates approximately at $1/3$ compared to $1/2$ for the SC. 
Next, the quadratic growth before the peak persists for longer times in higher-order GSCs, and as a result, they reach the peak at a later time compared to the SC. E.g., the peak in the ensemble-averaged  SC 
$\braket{\mathcal{C}^N(v)}$ is reached around $v \approx 0.67$, while the peak in $\braket{\mathcal{C}^N_{2}(v)}$ is reached around $v \approx 0.71$. 
 
Furthermore, the peaks in GSCs for $m \geq 2$ are more prominent compared to the SC. Since the peak in the SC is usually attributed to the chaotic nature of the energy spectrum, the GSCs of order $m \geq 2$ can be better indicators of chaos in this regard, as the peaks in these GSCs are sharper compared to the SC. It will be interesting to check whether this conclusion remains true for realistic quantum many-body systems that show an integrable to chaotic transition \cite{toappear}.

\section{Generalised spread complexity, and the $u$-Loschmidt echo for $su(2)$ Lie algebra}\label{su2}

In the previous section, we obtained analytical expressions of GSCs in the 
 continuum limit.   Another way one can proceed analytically is by considering
a quantum system for which the Hamiltonian is an element of some known Lie algebra.  
We start by considering such an example of computation where it is possible to obtain expressions for all orders of GSC analytically. 
Specifically, we consider the case where the Hamiltonian is an element of $su(2)$ Lie algebra, i.e., we assume
	\begin{equation}\label{su2H}
		H= \alpha (J_{+}+J_{-})+\gamma J_0 +\delta \mathbf{I}~,
	\end{equation}
where the operators $J_{i}$ are the generators of the $su(2)$ algebra, and hence, satisfy the usual commutation relation 
	\begin{equation}
		[J_0, J_{\pm}]= \pm J_{\pm}~,~~ [J_{+}, J_{-}]=2 J_{0}~. 
	\end{equation}
These generators act on the spin-$j$ representation in the usual way, where  $J_{+}$ and $J_{-}$ are raising and lowering operators.  
$\alpha$, $\gamma$, and $\delta$ are three constants, all of which we assume to be positive here. For the sudden quench setup mentioned in section \ref{statO}, the above Hamiltonian can be taken as the post-quench Hamiltonian, so that the time evolution after the quench at $t>0$ of an initial state (which can be taken as an eigenstate of the pre-quench Hamiltonian) is generated by it (see also the discussion towards the end of this section). 
	
In \cite{Balasubramanian:2022tpr}, the  SC of a time-evolved state 
generated by a Hamiltonian which is an element of the $su(2)$ algebra was obtained,  and the associated Krylov basis was constructed.
The Krylov basis elements are shown to be the spin-$j$ representation of the algebra, i.e., $\ket{K_n}=\ket{j, -j+n}$, and the expansion coefficients of the time-evolved state in this basis, obtained by using the standard decomposition formula for the $SU(2)$ group,  are given by
\begin{equation}
\phi_n(t)=\bigg(\frac{\Gamma (2j+1)}{n! \Gamma (2j-n+1)}\bigg)^{1/2}~e^{-i\delta t}\mathcal{A}(t)^n e^{-j\mathcal{B}(t)}~,
\end{equation} 
where the expressions for the time-dependent functions are 
	\begin{align}
		\mathcal{A}(t)=2\alpha ~\Bigg(i \Delta \cot \Big(\frac{\Delta}{2}t\Big) - \gamma \Bigg)^{-1}~,\\
		\mathcal{B}(t)= -2 \log \Bigg[\cos \Big(\frac{\Delta}{2}t\Big) + i \frac{\gamma}{\Delta} \sin \Big(\frac{\Delta}{2}t\Big)  \Bigg]~.
	\end{align}
Here we have denoted $\Delta=\sqrt{4\alpha^2 + \gamma^2}$. The simplified expression for the SC, as obtained in \cite{Balasubramanian:2022tpr}, is the following
	\begin{equation}
		\mathcal{C}(t)=\frac{8 \alpha ^2 j }{\Delta^2} \sin^2 \Big(\frac{\Delta}{2}t\Big)~. 
	\end{equation}
To obtain the expressions for higher-order GSCs, we first compute $G(\eta,t)$ defined in eq. \eqref{Generating}, which in this case, can be evaluated to be 
	\begin{equation}\label{Gsu2}
		\begin{split}
			G(\eta, t) = \bigg(1+\big(e^{\eta}-1\big)\frac{4\alpha^2}{\Delta^2} \sin^2 \Big(\frac{\Delta}{2}t\Big)\bigg)^{2j}~\\
			=  \bigg(1+\frac{1}{2j} \big(e^{\eta}-1\big)\mathcal{C} (t)\bigg)^{2j}~. 
		\end{split}
	\end{equation}
In the second line, we have written $G(\eta, t)$ in terms of the SC. For fixed values of $\eta$,  $G(\eta, t)$ is an oscillating function of time, whereas, for fixed values of $t$, this is an exponentially growing function of $\eta$.  
 
Now using the formula given in \eqref{generating}, we get the GSC of any order by taking the derivative of $G(\eta,t)$ at $\eta=0$. E.g., the GSC of order  $m=2$ is given by 
	\begin{equation}
		\begin{split}
		\mathcal{C}_2(t)= \frac{8j\alpha^2}{\Delta^4} \sin^2 \Big(\frac{\Delta}{2}t\Big) \Bigg (\Delta^2 - 
		4\alpha^2 (1 - 2 j)\sin^2\Big (\frac{\Delta}{2}t\Big)\Bigg)~\\
		= \mathcal{C}(t) + \Big(1-\frac{1}{2j}\Big) \mathcal{C}(t)^2~.
	\end{split}
	\end{equation}
From these expressions, the variance of the spreading defined in eq. \eqref{variance} can be calculated to be, 
	\begin{equation}
		\begin{split}
		(\Delta n(t))^2  =  \frac{8j\alpha^2}{\Delta^4}  \sin^2 \Big(\frac{\Delta}{2}t\Big) ~ \Bigg(\Delta^2 -4\alpha^2  \sin^2 \Big(\frac{\Delta}{2}t\Big)\Bigg)~
		= \mathcal{C}(t) -  \frac{\mathcal{C}(t)^2}{2j} ~. 
    \end{split}
	\end{equation}
Notice that, the fact   $\Delta^2 > 4\alpha^2  \sin^2 \Big(\frac{\Delta}{2}t\Big)$ ensures that the variance of the spreading always remains positive during the time evolution, though its value can be zero at certain instants of time. The time instances when the variance goes to zero coincide with zeros of the SC, i.e., for these values of time after the start of the evolution, both the mean and variance of the distribution of the spreading vanish.  
	
\begin{figure}[h!]
		\centering
		\includegraphics[width=3in,height=2.2in]{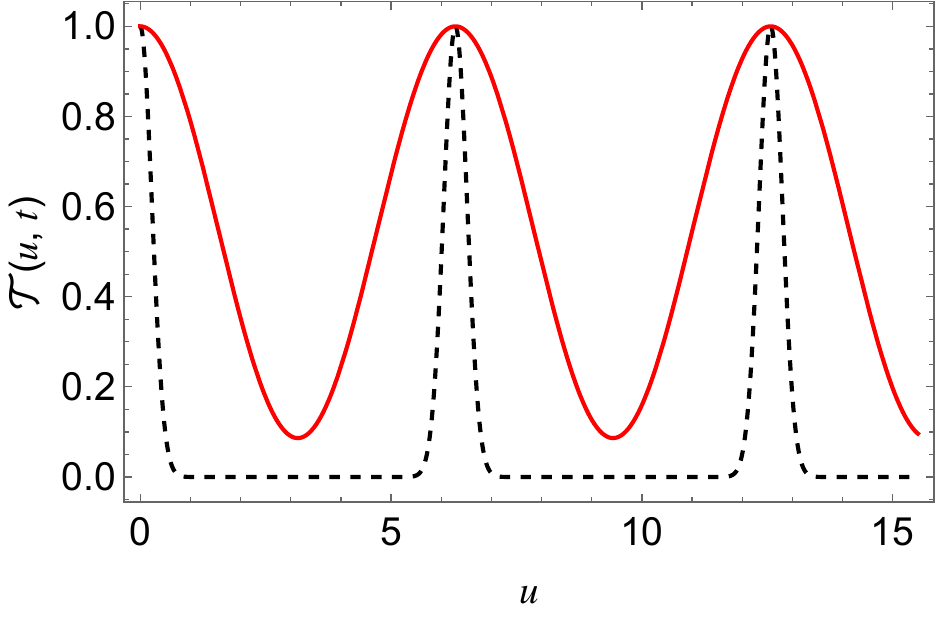}
		\caption{Plots of the transition probability of the $u$ evolution (i.e., the $u$-Loschmidt echo) for fixed value of $t=1$ and for two different values of $j$, $j=1/2$ (red) and $j=10$ (dashed-black). We have taken  $\alpha=1, \gamma=1$ for both cases.  }
		\label{fig:uLE}
	\end{figure}

For this example, we can compute the expression for the transition probability of the $u$-evolution (the $u$-Loschmidt echo), $\mathcal{T}(u,t) = |\chi_{\mck}(u,t)|^2 $, as well as the exact expression for the PDF for the statistics of the spreading operator. 
	Using the expression for the generating function in eq. \eqref{Gsu2}, we see that $\mathcal{T}(u,t) $  is given by, 
	\begin{equation}\label{LEsu2}
		\mathcal{T}(u,t) 	=  \bigg(1+  \frac{1}{j}  \sin^2 (u/2) ~ \mathcal{C}(t)\Bigg[\frac{\mathcal{C}(t)}{2j} -1\Bigg]\bigg)^{2j}~. 
	\end{equation}
For a fixed value of $u$, the instances when the SC (i.e., the mean of the distribution in eq. \eqref{SOsta}), as well as the variance, go to zero are precisely the values of time when the transition probability of $u$-evolution returns to its initial value. A plot of the function for a fixed value of $t$ and different values of $j$ is shown in Fig. \ref{fig:uLE}. 
Even though the overall behaviour of the transition probability shows complete revivals,  clearly, for large values of $j$, it decays rapidly and remains zero for a long period of $u$ and then again grows swiftly to its initial value. 

Taking the inverse Fourier transform of the characteristic function ($\chi_{\mck} (u,t) =G(-iu,t) $) and using the expression in \eqref{Gsu2}, we obtain the probability distribution of the measurements of the spreading operator. E.g., when $j=1/2$, the PDF associated with a measurement outcome of $n=x$ is given by 
	\begin{equation}
		P(x,t) = \delta(x) \Bigg(1-\frac{4 \alpha^2}{\Delta^2} \sin^2 \Big(\frac{\Delta}{2}t\Big)\Bigg) +\delta(x-1) \Bigg(\frac{4 \alpha^2}{\Delta^2} \sin^2 \Big(\frac{\Delta}{2}t\Big)\Bigg)~.
	\end{equation}
The two delta functions are due to the fact that for $j=1/2$, the Krylov space is two-dimensional. This expression can also be directly obtained from the definition in eq. \eqref{SOsta}, and using the expressions for the transition probabilities.
	
	
By following a similar procedure as in this section, we can also obtain the expressions for the GSCs for cases when the Hamiltonian is an element of other Lie algebras, such as $su(1,1)$,  or the Heisenberg-Weyl algebra. In these cases as well, it is possible to obtain a direct relation between the SC and the function $G(\eta,t)$, similar to the one in eq. \eqref{Gsu2}.\footnote{See section \ref{unitary} for such a relation for the case of the $su(1,1)$ Lie algebra.} The importance of considering such Lie algebraic Hamiltonians is that they are used quite frequently to model different quantum mechanical systems with some underlying symmetries.  E.g.,  $su(1,1)$ is one of the algebras that often appear in quantum quench setups, since,  when quenches of Hamiltonians which can be written in a non-interacting form in terms of a set of bosonic creation and annihilation operators and are quadratic functions of these operators are considered, in many cases, it is possible to rewrite the post-quench Hamiltonian as a linear combination of operators which are the three generators of the $su(1,1)$ Lie algebra. Prominent examples of such systems include sudden quenches of a chain of harmonic oscillators with nearest neighbour interaction \cite{spread2} and the Lipkin-Meshkov-Glick model in the thermodynamic limit \cite{spread1}.  In the next section, we use the $su(1,1)$ algebra in a different context to study the behaviour of GSC under a unitary transformation of the Hamiltonian.  

\section{Unitary transformations and the generalised spread complexity}\label{unitary}
The purpose of the present section is twofold.   We first consider the effect of a time-independent unitary transformation of the Hamiltonian on the Krylov basis, the GSC, and hence on the statistics of wavefunction spreading.  Specifically,  we discuss that the GSC and the characteristic function $\chi_{\mck} (u,t) $ remain invariant under such a transformation. Next, we use this generic property for a specific class of Hamiltonians belonging to the $su(1,1)$ algebra and consider the effect of a unitary transformation by the displacement operator of the group.  
We show that SC associated with two suitably chosen reference states can be directly related to each other, and we use this relation to provide a possible geometrical interpretation of SC when the Hamiltonian is an element of some Lie algebra, following ref. \cite{Caputa1}, which discussed a similar interpretation for the Krylov operator complexity. 

Consider first the effect of a time-independent unitary transformation $U$ of the Hamiltonian on the Krylov basis and the statistics of wavefunction spreading. Denoting the  new transformed Hamiltonian $\tilde{H} = U^{\dagger} H U $,  and assuming that the new reference state is related to the previous one by $\ket{\tilde{\psi}_0}= U^\dagger \ket{\psi_0}$, 
it is easy to check from the action of the Hamiltonian on the Krylov basis elements that,  the new Krylov basis $\ket{\tilde{K}_n}$ are related to the old one through the relation $\ket{\tilde{K}_n} =  U^{\dagger} \ket{K_n} $, while the Lanczos coefficients remain unchanged.   
Therefore, the expansion coefficients ($\phi_n(t)$) of the time-evolved state in the Krylov basis, and hence the SC as well as GSC of all orders, remain unchanged under the time-independent unitary transformation \cite{Gill:2023umm, Beetar:2023mfn}.  Alternatively, this can be understood from the fact that the spreading operator transforms as $\tilde{\mathcal{K}_m} = U^{\dagger} \mathcal{K}_m U$, hence GSCs are invariant under the unitary transformation, $\tilde{\mathcal{C}}_m(t)=\mathcal{C}_m(t)$. Since the GSCs are invariant under this transformation, the moments of the characteristic function, and hence, the characteristic function itself, remain invariant under this transformation. 
In summary, the GSC associated with time-evolved state $\ket{\Psi(t)}= e^{-i H t} \ket{\psi_0}$  (obtained from the initial state $\ket{\psi_0}$ through an evolution by $H$),  is the same as GSC associated with the time evolved state $\ket{\tPsi (t)}= U^{\dagger} \ket{\Psi(t)} =  e^{-i \tilde{H} t} \ket{\tilde{\psi}_0}$ (obtained from the initial state $\ket{\tilde{\psi}_0}= U^\dagger \ket{\psi_0}$ through an evolution by $\tilde{H}$). 

Next consider the  following Hamiltonian belonging to the $su(1,1)$  Lie algebra,  
	\begin{equation}\label{su11Hm}
		H= 2\lambda (e^{i \beta} K_{+}+e^{-i\beta}K_{-})+2\omega  K_0 ~,
	\end{equation}
where $\lambda$, $\beta$, and $\omega$ are three real constants which we assume to be positive.  The three generators satisfy the standard commutation relations for this algebra:
	\begin{equation}
		[K_0, K_{\pm}]= \pm K_{\pm}~,~~ [K_{-}, K_{+}]=2 K_{0}~. 
	\end{equation}
The Hamiltonian of the form given in \eqref{su11Hm} appears in quantum optics in various contexts, one such example being the degenerate parametric amplifier \cite{Wodkiewicz, Gerry1, Gerry2}. Assuming that the initial state $\ket{\psi_0}$ (i.e., the first state of the Krylov basis) is the state $\ket{h,n=0}$ of the discrete series representation of the algebra (where $h$ is the Bargmann index of the algebra), the Krylov basis associated with the above Hamiltonian (with $\beta=0$) and the SC of the time-evolved state $\ket{\Psi(t)}=e^{-i H_{\beta=0} t}\ket{\psi_0}$  had been obtained in \cite{Balasubramanian:2022tpr}. 
		
To understand the consequence of the unitary invariance discussed above for this Hamiltonian, we perform a unitary transformation of it using  the following operator
	\begin{equation}\label{dispacemnt1}
		\mathcal{D} (\txi) = \exp \Big(\txi K_{+} - \bar{\txi}  K_{-}\Big) ~.
	\end{equation}
This is just the displacement operator associated with the $su(1,1)$ algebra. Here $\txi$ is a complex variable which can be written in terms of two real  parameters  $\theta$ and $\phi$ as 
	\begin{equation}
		\txi = -\frac{\theta}{2} e^{-i \phi}~,~~-\infty < \theta < \infty~, ~ 0 \leq \phi \leq 2\pi~. 
	\end{equation}
The final simplified expression for the Hamiltonian after this transformation is given by  \cite{Gerry2}
	\begin{equation}
		\tilde{H} = \mathcal{D}^{\dagger} (\txi)  H \mathcal{D} (\txi) = \mathbb{A}_0 (\theta, \phi) K_0 + \mathbb{A}_{+} (\theta, \phi)  K_{+} 
		+ \mathbb{A}_{-}(\theta, \phi)  K_{-}~,
	\end{equation}
where the new coefficients of the three generators are
	\begin{eqnarray}
		\mathbb{A}_0 (\theta, \phi)  =  2 \omega \cosh \theta  + \frac{4 \lambda}{(1- |\tz|^2)}  \Big(\tz e^{i \beta} + \bar{\tz} e^{-i\beta}\Big)~,\label{A0}\\
		\mathbb{A}_+ (\theta, \phi)  =  - \omega e^{-i \phi} \sinh \theta + \frac{2 \lambda}{(1- |\tz|^2)}   \Big(\tz^2 e^{i \beta} + e^{-i\beta}\Big)~,\\
		\text{and}~~ \mathbb{A}_- (\theta, \phi)  = \bar{\mathbb{A}}_+ (\theta, \phi)~. 
	\end{eqnarray}
Here the complex variable $\tz$ is related to the parameter $\theta$ and $\phi$ through the relation
	\begin{equation}
		\tz = - \tanh \big(\theta/2\big) e^{-i\phi}~,
	\end{equation}
and a bar denotes a complex conjugation.
	
The strategy we now follow is to find a particular set of values of the parameters $\theta$ and $\phi$ such that the coefficient $\mathbb{A}_0(\theta,\phi)$ vanishes for that choice.  From Eq. \eqref{A0}, it is easy to see that when we take $\phi=\beta$ and $\theta  = \tanh^{-1} (\omega/ 2\lambda)$, the coefficient $\mathbb{A}_0$ reduces to zero, and the final form for the transformed  Hamiltonian simplifies to 
	\begin{equation}\label{Htilde}
		\tilde{H} =  \mathbb{A}_{+} K_{+} 
		+ \mathbb{A}_{-} K_{-}=	\Delta e^{-i\beta} K_{+} + \Delta e^{i\beta} K_{-}~,
	\end{equation}
where we have defined the constant $\Delta= \sqrt{4\lambda^2-\omega^2}$.\footnote{This $\Delta$ is different from the one defined in  section \ref{su2}.}  Note that, we have assumed that  the condition $2\lambda > \omega$ is satisfied (i.e., $\Delta$ is a real constant) when deriving \eqref{Htilde}. 
Furthermore, to connect with the results presented in \cite{Balasubramanian:2022tpr},
sometimes we shall put $\beta=0$ in the followings. 
	
From the unitary invariance discussed at the beginning of this section, the GSC of the time-evolved state $\ket{\Psi(t)}= e^{-i H t} \ket{h,0}$ is the same as the state $\ket{\tPsi (t)}=  
	e^{-i \tilde{H} t} \ket{\txi, h,0}$ (i.e., $\tilde{\mathcal{C}}_m(t)=\mathcal{C}_m(t)$), where $\ket{\txi, h,0}$ is the initial state of the $\tilde{H}$ evolution ($\ket{\tilde{\psi}_0}$). Also note that, since the unitary operator $\mathcal{D} (\txi)$ is the displacement operator of the algebra, the state $ \ket{\txi, h,0}$ is actually a Perelomov coherent state associated with the $su(1,1)$ Lie algebra. 

We now compare the Lanczos coefficients, Krylov basis, and the GSC in two different time-evolution scenarios (the time evolution in both cases is generated by the unitary transformed Hamiltonian $\tilde{H}$). The reason for doing so will be clear from the relations we shall derive later. The two cases are as follows:
 \begin{itemize}
     \item \textbf{Case-I}: The initial state before the start of the evolution is the transformed state $\ket{\tilde{\psi}_0}=\ket{\txi, h,0}$. The Lanczos coefficients and the Krylov basis are receptively given by \cite{Balasubramanian:2022tpr}:
     \begin{equation}
		\tilde{a}_n = 2 \omega (h+n)~,~~\tilde{b}_n = 2\lambda \sqrt{n (2h+n-1)}~,
	\end{equation}
 and $\ket{\tilde{K}_n}= \mathcal{D}^{\dagger} (\txi) \ket{h,n}$.
 \item \textbf{Case-II}\footnote{We denote quantities associated with case-II with a prime for convenience. Note that though here the evolution is generated by the same Hamiltonian $\tilde{H}$ as in case-I, the initial state is different, so that the Lanczos coefficients and the GSC are different from case-I.}: The  initial state is the original state $\ket{\psi^{\prime}_0}=\ket{\psi_0}=\ket{ h,0}$. Then the Lanczos coefficients and Krylov basis are given by 
 \begin{equation}
		a^{\prime}_n = 0~, ~~ b^{\prime}_n = \Delta  \sqrt{n (2h+n-1)}~,
	\end{equation}
and $\ket{K^{\prime}_n} = \ket{h,n}$. Note that the Lanczos coefficients  $a^{\prime}_n$ associated with $\tilde{H}$ and the initial state  $\ket{h,0}$ (i.e., case-II) are zero, i.e., the unitary transformation  \eqref{dispacemnt1} performed essentially maps the generic Hamiltonian spreading problem to a situation where the first set of Lanczos coefficients is identically zero.\footnote{When the initial state is assumed to be the same; otherwise, the Lanczos coefficients remain unchanged. } 
 \end{itemize}

To obtain the time-evolved state in both cases, we note that, the time evolution operator generated by the transformed  Hamiltonian, 
	\begin{equation}
		U(t)=  \exp \Big[-i t \tilde{H} \Big] =    \exp \Big[-i t ~ \Big(\mathbb{A}_{+}  K_{+} 
		+ \mathbb{A}_{-} K_{-}\Big)\Big] ~
	\end{equation}
has  the form of the displacement operator  $ \mathcal{D} (\xi)$ with $\xi= -it \mathbb{A}_+ = -it \Delta e^{-i\beta} $.  
Therefore, the time evolved state $\ket{\Psi ^{\prime} (t)}= e^{-i \tilde{H} t} \ket{h,0}$ is a Perelomov coherent state of the $su(1,1)$ Lie algebra.  Using this identification, it is straightforward to obtain the SC of the time-evolved state following the method proposed in \cite{Caputa1}. 

To obtain the wavefunctions $\phi^{\prime}_n (t)$ in case-II, we parametrise the time-evolved coherent state by the complex variable 
	\begin{equation}
		z =  \tanh \big(\theta^\prime/2\big) e^{i\phi^\prime}~,
	\end{equation}
where $\xi= \frac{1}{2} \theta^\prime e^{i\phi^\prime}$. The variable $z$ parametrises a unit disk with $|z|<1$. Notice that here we have $\phi^{\prime} = \frac{\pi}{2} - \beta$, and $\theta^{\prime}= -2t\Delta$.  Therefore,   $z= -  i \tanh (t\Delta)  e^{-i\beta}$.  The wavefunctions ($\phi_n^\prime (t)$) can be obtained from the expansion of the state $\ket{\Psi ^{\prime} (t)}$  in the Krylov basis $\ket{K^{\prime}_n} = \ket{h,n}$. These  are given by (with $\beta=0$)
	\begin{equation}
		\phi_n^\prime (t)=  N_n (1-|z|^2) ^h  z^n = (-i)^n(1-\tanh (t\Delta )^2) ^ h  \tanh (t \Delta )^n~, 
	\end{equation}
where $N_n=\sqrt{\frac{\Gamma (2h+n)}{n! \Gamma (2h)}}$.  Using these expressions for the wavefunctions, we can easily obtain the 
SC to be $\mathcal{C}^{\prime}(t)= 2h \sinh^2 (t \Delta ) = 2h \sinh^2 (\sqrt{4\lambda^2 - \omega^2} t)$.  On the other hand, the formula for the SC in case-I is given by  $\tilde{\mathcal{C}}(t)=  \frac{8\lambda^2 h}{\Delta^2} \sinh^2 (\sqrt{4\lambda^2 - \omega^2} t)$. 
Therefore, one can straightforwardly relate the SC in the two cases under consideration and see that they are, in fact, proportional to each other,
\begin{equation}\label{com12}
		\tilde{\mathcal{C}}(t) =\mathcal{C}(t)=  \frac{4\lambda^2 }{\Delta^2} \mathcal{C}^{\prime}(t)~.
	\end{equation}
Due to the unitary invariance of the SC, we can use the identification that $\mathcal{C}(t)= \tilde{\mathcal{C}}(t)$. 

\textbf{An interpretation of the SC ($\mathcal{C}(t)$).} From the relation between $\mathcal{C}(t)$ and $\mathcal{C}^{\prime}(t)$ for case-I and case-II derived in \eqref{com12} it is possible to provide a geometric interpretation of the SC ($\mathcal{C}(t)$) following \cite{Caputa1}, where such an interpretation was established for Krylov operator complexity when the Liouvillian operator is a linear combination of raising and lowering ladder operators of some well-known Lie algebras (such as the $sl (2, R)$, $su(2)$ and Heisenberg-Weyl algebra).  The connection obtained in \cite{Caputa1} suggests that when the Liouvillian operator is taken as a linear combination of raising and lowering ladder operators of a Lie algebra,  the Krylov complexity is proportional to the volume bounded by the geodesic of the information geometry corresponding to the Lie group under consideration, where the exact parametrisation of geodesic can be obtained by comparing the time evolved state (in the space of operators) with the Perelomov coherent states of the Lie group. 

However, due to the presence of the additional operator $K_0$ in the Hamiltonian \eqref{su11Hm}, it is not straightforward to extend such an interpretation directly to the case of SC.  Nevertheless, we can use the relation $\mathcal{C}(t) =  \frac{4\lambda^2 }{\Delta^2} \mathcal{C}^{\prime}(t)$ obtained above to 
	provide a geometrical interpretation of the SC of a time evolved state generated by \eqref{su11Hm}, since the quantity $\mathcal{C}^{\prime}(t)$ has a similar geometric interpretation as the Krylov operator complexity. Since, a relation such as the one in \eqref{com12} can also be obtained when the Hamiltonian is an element of Lie algebras other than $su(1,1)$, such as the $su(2)$ and Heisenberg-Weyl algebras, we conclude that for such Hamiltonians, the SC is proportional to the volume bounded by the geodesic on the information geometry of the associated Lie group. However,  the proportionality constant is different from a similar relation between the Krylov operator complexity and this volume. \footnote{See also \cite{Seetharaman:2024ket} for an alternative geometrical interpretation of the SC for a qubit system.} 

\textbf{Generating function and higher order complexities.} Next, we derive a relation between the generating function and the SC for these two cases.  For case-II,  the relation between  the function $G(\eta,t)$ introduced in Eq. \eqref{Generating} (defined suitably for this case) and the SC is  
	\begin{equation}\label{generating2}
		\begin{split}
			G^{\prime}(\eta,t)= \bigg(1-\big(e^{\eta}-1\big) \sinh^2 \big(t \Delta  \big)\bigg)^{-2h}~
			=  \bigg(1-\frac{1}{2h} \big(e^{\eta}-1\big)\mathcal{C}^{\prime} (t)\bigg)^{-2h}~,
		\end{split} 
	\end{equation}
while in case I this  relation can be written as 
	\begin{equation}\label{generating1}
		\begin{split}
			G(\eta, t) = \bigg(1-\big(e^{\eta}-1\big)\frac{4\lambda^2}{\Delta^2} \sinh^2 \big(t \Delta \big)\bigg)^{-2h}\\
			=  \bigg(1-\frac{1}{2h} \big(e^{\eta}-1\big)\mathcal{C} (t)\bigg)^{-2h}~.
		\end{split}
	\end{equation}
These two relations clearly show that, though the expressions for the generating function and the SC are different, the relation between the two of them is the same for the original Hamiltonian and the unitary transformed Hamiltonian (when both are elements of the same Lie algebra), irrespective of the initial state. Furthermore, one can see that the relation between $\mathcal{C}_m(t)$s of higher orders (i.e., $m \geq 2$) and the respective SC are the same in both cases. 
In fact, the structure of the generating function above shows that the GSC of order $m$  can be expressed as a polynomial function of the SC of degree $m$. As an example, from eq. \eqref{generating1} it can be seen that the GSC of $m=2$ is a quadratic polynomial  of the SC, 
\begin{equation}
		\mathcal{C}_2(t)=\Big(1+\frac{1}{2h} \Big) \mathcal{C}^2(t)+ \mathcal{C}(t) ~,
\end{equation}
	so that the variance of the spreading is given by $	(\Delta n(t))^2   = \frac{\mathcal{C}^2(t)}{2h}+ \mathcal{C}(t)$. 
Using \eqref{generating2}, one can now check that the relation between $	\mathcal{C}^{\prime}_2(t)$ and $\mathcal{C}^{\prime}(t)$ is exactly the same as above. 
However, in general, it is difficult to obtain a simple relation between $	\mathcal{C}^{\prime}_2(t)$  and $	\mathcal{C}_2(t)$  (or between higher order GSCs).  
		
Before concluding this section, we note the following two points. Firstly, though the GSC of any order is polynomially related to the SC when the Hamiltonian is an element of the previously mentioned Lie algebras, at this point, it is not clear to us whether it is also possible to provide any geometrical interpretation of the higher-order GSCs as well. Secondly, one can define a form of complexity from the information geometry associated with a classical or quantum system. For a classical system, the information geometry is essentially the Fisher-Rao metric obtained from the probability distribution function associated with the parameters describing the system, and for a quantum system in a pure state, this is the Fubini-Study metric. This notion of complexity, known as the information geometric complexity,  is then obtained by integrating the volume bounded by a geodesic in this geometry up to 
 some value of the parameter that parameterises the geodesic, and can also be the time itself \cite{cafaro2010reexamination, felice2018information}. Though this picture of complexity, for quantum systems, is not defined with respect to a specific basis, it would certainly be interesting to see whether this can be related to the SC studied in this paper or whether one can generalise the geometric interpretation of SC for Lie algebras we have discussed in this section to unitary evolution in a generic quantum system to provide a definition of SC from the perspective of the geometry of quantum states. 
		
\section{A bound on the total change of generalised complexity   }\label{bounds}
In this section we derive an upper bound on the change of the  GSC up to a finite time (denoted here as $\tau$)  under a Hamiltonian evolution.   
The primary goal of obtaining an upper bound on the change of an information-theoretic quantity is to quantify the energetic cost associated with the change of `information'  of an initial state under a unitary Hamiltonian evolution.
Since, the usual quantifier of the information associated with a generic quantum state, the von Neumann entropy, does not change under a unitary evolution, to quantify the change of information of a state, one possible measure can be the Shannon entropy, which does change under Hamiltonian evolution \cite{Deffner}.  A natural question, therefore, one can ask in this context is what is the total change of `complexity' of a state under a Hamiltonian evolution up to a finite time?  
 Here, we use the definition of complexity associated with the spreading of a pure state discussed so far in this paper to quantify such a change. 

The derivation provided here follows a similar derivation of an upper bound on the change of the Shannon information under a Hamiltonian evolution presented recently in \cite{Deffner}, where such a bound was obtained in terms of the norm of the Hamiltonian.\footnote{For the derivation of the bound on the change of the Shannon information \cite{Deffner},  Hamiltonian can be time-dependent in general. In this paper, we assume that the Hamiltonian is not an explicit function of time.}   Since, both the Shannon entropy and the SC of a time-evolved state essentially depend on a set of 
 probabilities computed with respect to a specific basis, while in the former case, this basis can be any arbitrary basis, in the latter case this basis is chosen to the Krylov basis by definition of the SC,  it is natural to ask whether one can obtain a useful bound on the change of the SC and see how these transition probabilities in the Krylov basis bounds the spreading of an initial state with time. As we shall see, the bound on the total change of the SC is essentially determined by the weighted average of the absolute value of these probability amplitudes and the operator norm of the Hamiltonian.  
Furthermore, as we demonstrate towards the end of this section, one can, in fact, consider the evolution of the Shannon entropy in the Krylov basis and obtain a bound on its change. We believe these bounds will provide a better understanding of the nature of the quantum dynamics in the Krylov basis. 

To start, let us denote the probability of obtaining the time evolved state $\ket{\Psi(t)}$ in the $n$-th Krylov basis by $p_n(t)$, i.e.,  $p_n(t)= | \braket{K_n|\Psi(t)} |^2 = |\phi_n(t)|^2$. Taking a time derivative of these functions and using the  Schrodinger equation, we can derive the following inequality,
	\begin{equation}
		\dot{p}_n(t) \leq 2~\big|\bra{K_n} H \ket{\Psi(t)} \braket{\Psi(t)|K_n}\big|~.
	\end{equation}
	Using the Holder inequality \cite{Baumgartner}, it is possible to further modify the above inequality as \cite{Deffner}
	\begin{equation}\label{dotpineq}
		\dot{p}_n(t) \leq  2 \sqrt{p_n(t)}  ~||H|| ~,
	\end{equation}
where $||H||$ denotes the operator norm of the Hamiltonian,\footnote{For a matrix representation of the Hamiltonian in a suitable basis, its operator norm is given by the square root of the largest eigenvalue of the matrix $H^{T}H$. Here $H^{T}H=H^2$,  since the Hamiltonians we consider in this paper are Hermitian matrices.} and $\sqrt{p_n(t)}= |\phi_n(t)|$. Explicitly, in terms of the Hamiltonian and the energy eigenvalues, we have 
\begin{equation}
		\begin{split}
			\dot{p}_n(t) \leq  2  ~||H|| ~ \big|\braket{\psi_0|~\mathcal{P}_n(H)e^{-itH} ~|\psi_0}\big|~\\
			= 2  ~||H||  \sum_a  \mathcal{P}_n (E_a) ~\rho_0(E_a) ~e^{-it E_a}~.
		\end{split}
\end{equation}
	Alternatively, introducing the parameters $\vartheta_n (t)$ through $\cos \vartheta_n(t)= \sqrt{p_n(t)}= |\phi_n(t)|$, we can rewrite \eqref{dotpineq} as
	\begin{equation}
		\dot{\vartheta}_n (t)  \geq - ||H||/ \sqrt{1- \cos^2\vartheta_n(t)} ~. 
	\end{equation}
	Note that the condition $\vartheta_n= \pi/2$ indicates that the $n$-th element $\ket{K_n}$ of the Krylov basis is orthogonal to the time-evolved state.
	
Using the expressions for the change in the transition probabilities in the Krylov basis, we can now derive an upper bound on the total change of the SC after a finite time as follows. By using the triangle inequality and the definition of the SC, we have \footnote{Here the symbol $	|\Delta \mathcal{C}| $ indicates the magnitude of the change of SC in a finite time. It should not be confused with the variance of the spreading in eq. \eqref{variance}. }
	\begin{equation}
		|\Delta \mathcal{C}| \leq \int_{0}^{\tau} \text{d}t ~ |\dot{\mathcal{C}}(t)| = \int_{0}^{\tau} \text{d}t ~\big |\sum_n n \dot{p}_n(t)\big|~.
	\end{equation}
	Substituting the inequality derived in \eqref{dotpineq}, we obtain the following bound on the change of the SC 
	\begin{equation}\label{SCbound}
		|\Delta \mathcal{C}|  \leq 2 ||H||  \int_{0}^{\tau} \text{d}t ~ \mathcal{F}(t)~= 2 ||H|| \tau ~\bar{\mathcal{F}}(\tau)~ ,
	\end{equation}
	where we have defined the function
	\begin{equation}
		\mathcal{F}(t)= \sum_n n \sqrt{p_n(t)} =  \sum_n n \cos \vartheta_n(t) ~,
	\end{equation}
	which is always positive, and in the second expression, we have written it in terms of its time average. By following an entirely similar procedure, we can also derive similar bounds on the GSC 
	\begin{equation}
		|\Delta \mathcal{C}_m| \leq  2 ||H||  \int_{0}^{\tau} \text{d}t ~ \mathcal{F}_m(t)~,~\text{with}~~\mathcal{F}_m(t)= \sum_n n^m \sqrt{p_n(t)}~.
	\end{equation}
For a finite-dimensional Hamiltonian, we can see that the bound of the GSC is invariant under a unitary transformation,\footnote{One example could be the transformation 
considered in the previous section, i.e., case I.}
since the singular values of the Hamiltonian matrix are invariant under a unitary transformation, and the probability amplitudes, as we have discussed in section \ref{unitary}, are also invariant under a unitary transformation.  One further point we note as well is that, since the moments of the characteristic function of the $u$-evolution are directly related to the GSC (see eq. \eqref{GSC=Mm}), the bounds on the change of the GSC can be translated to possible bounds on the change of these moments. 
	
For a Hamiltonian and an initial state for which the associated Krylov subspace is two-dimensional, the bound on the change of the SC is related directly to the SC itself.
In such a case, $\mathcal{C}(t)= \mathcal{C}_m(t)= p_1(t)$, and hence, $\mathcal{F}(t)=\mathcal{F}_m(t)= \sqrt{\mathcal{C}(t)}$. Therefore, the bound on the change of the SC is given by
	\begin{equation}\label{boundd=2}
		|\Delta \mathcal{C}|  \leq 2 |E_0|  \int_{0}^{\tau} \text{d}t  \sqrt{\mathcal{C}(t)}~,~\text{for}~~L=2~,
	\end{equation}
where $|E_0|$ denotes the absolute value of the largest eigenvalue of the Hamiltonian.  

To check this bound explicitly, we consider a system with two-dimensional 
Hilbert space, and the time evolution of an initial state on the Bloch sphere. The resulting Krylov basis is also two-dimensional. 
Parametrising the initial location of the state on the Bloch sphere by the angles $0 \leq \theta_0 \leq \pi$, and $0 \leq \phi_0 \leq 2\pi$, one can derive the expression for the SC to be \cite{Aguilar-Gutierrez:2023nyk}:  $\mathcal{C}(t)= \sin^2 (\theta_0) \sin^2 (\Delta E t/2)$. 
Here, $\Delta E =E_0-E_1$ is the difference between the two energy eigenvalues of the system. We assume that both energies are positive, and 
$E_0>E_1$, so that $\Delta E$ is also positive.  Using this expression, we can derive the exact value of the change of the SC (the left-hand side 
of \eqref{boundd=2}), as well as the bound in the right-hand side of eq. \eqref{boundd=2}. The analytical expression for the ratio ($r$) of these two quantities  is given by 
\begin{equation}
		r= \frac{\Delta E}{2 E_0} \cos^2 \bigg(\frac{\Delta E \tau}{4}\bigg) \sin(\theta_0)~. 
\end{equation}
Since $\Delta E< E_0$, it is easy to see that $r<1$ for any value of $\tau$. However, depending upon relative values of $\Delta E$ and $E_0$, the bound derived above may not be 
tight for some values of $\tau$, i.e., $r \ll 1$. In fact,  if we consider two Hamiltonians of this type, with different numerical values of the coefficients such that $\Delta E$ is the 
same for both of them and only the magnitudes of $E_0$ and $E_1$ are different, one can check that the ratio above provides a relatively tighter bound when $E_0$ is smaller. The bound above starts to become less useful as the operator norm of the two-level Hamiltonian (i.e., $E_0$) and the other eigenvalue ($E_1$) increase in magnitude. 
	
We conclude this section by noting the following points. 
1. The change of the  spread entropy, i.e., the Shannon entropy in the Krylov basis, has the following bound 
	\begin{equation}
		|\Delta \mathcal{S}_K| \leq 2 ||H||  \int_{0}^{\tau} \text{d}t ~ \mathcal{G}(t)~,~\text{with}~~\mathcal{G}(t)= - \sum_n \log p_n (t) \sqrt{p_n(t)}~.
	\end{equation}
This expression is similar to the one given in \cite{Deffner}; the only change is that the probability amplitudes here are computed with respect to the Krylov basis. 
	
	2. From the expression for the bound in the change of the SC derived in eq. \eqref{SCbound}, we see that it is dependent on the finite time average of the function $\mathcal{F}(t)= \sum_n n \sqrt{p_n(t)} = \sum_n n |\phi_n(t)|$. 
It is, in principle, possible to think of this quantity as a modification of the cost function associated with the spreading of a time-evolved wavefunction defined in eq. \eqref{cost}. As it turns out, it is easy to derive a 
	rather simple expression for the bound on the change of this modified cost under the Hamiltonian evolution. Following the same procedure as outlined above and using the inequality in \eqref{dotpineq}, we have the  expression for the bound on the change of the modified cost to be 
\begin{equation}
		|\Delta \mathcal{F}|  \leq   ||H|| ~ \tau ~ \Big| \sum_{n} n\Big|~.
\end{equation}
	This bound depends only on the operator norm of the Hamiltonian and always grows linearly with the final time. 
	For a finite-dimensional Krylov basis, one can easily evaluate the above bound.  For example, in the case when the Krylov basis is two-dimensional, this bound is just 
	$	|\Delta \mathcal{F}|  \leq  |E_0|  \tau $, i.e., the total change of this cost can not exceed a linear growth.
	For an infinite-dimensional Krylov basis, we need to implement a suitable regularisation procedure to evaluate the sum.  At this point, we note that, unlike the cost defined in \eqref{cost}, this modified cost may not be minimised in the Krylov basis for a finite time around the initial time $t=0$, and hence, it is not justified to call this a modified measure of the complexity of the wavefunction spreading without proving this explicitly. We hope to return to the problem of determining whether this modified cost function is also minimised up to a finite time when computed in the Krylov basis in future work.
	
	3. In this section, we have obtained an upper bound on the total change of  GSC up to a finite time in a unitary evolution in terms of the operator norm 
	of the Hamiltonian generating the evolution. In contrast,  previous works have studied the bound on the rate of the growth of the Krylov operator complexity (or the K-complexity), and its various related measures with time. 
	As an example,  It was shown in \cite{Hornedal:2022pkc}, that the growth rate of the K-complexity satisfies a bound determined by the first Lanczos coefficient, $b_1$ of the Liouvillian recursion, the variance of the K-complexity operator, which one can recast in a form similar to the Mandelstam-Tam bound. 
A similar bound was later established for the growth rate of the Krylov operator entropy in \cite{Fan:2022mdw}, in terms of $b_1$ and the variance of the Krylov entropy operator. 

\section{Summary and discussions} \label{summary}
To summarise, in this paper, we have defined and studied various properties of the statistics of the results of a measurement of the spreading operator $\mck$ in the Krylov basis, where the basis elements are constructed from an initial pure state and the Hamiltonian generating the subsequent time evolution through the Lanczos algorithm.	 The characteristic function of this distribution (obtained from the Fourier transformation of the probability distribution of the eigenvalues of $\mck$) can be interpreted as a correlation function, which essentially represents an overlap between the initial state and an evolved version of it, where this evolution is generated by the Heisenberg time-evolved spreading operator.  We show that the moments of this characteristic function are related to the various powers of the average spreading of the time-evolved state in the Krylov basis, which are known as the GSCs. Therefore, the GSCs contain all information about the statistics of the spreading of an initial state in the Krylov basis. 

In the rest of the paper, we have studied various properties of these GSCs for 
several Hamiltonians for which one can compute these quantities analytically. To reiterate briefly, in section \ref{propertiesGSC}, we obtain an analytical 
expression for the statistics when the Hamiltonian is an element of the 
$su(2)$ Lie algebra, and also provide a geometrical interpretation of the variance of the spreading. Another important property of the GSCs is that they are invariant under any time-independent unitary transformation of the Hamiltonian generating the time evolution.  In section \ref{unitary}, we use this property to show that the relation between the generating function (which is the analytical continuation of the characteristic function to imaginary values of the parameter, which is conjugate to the result of measurement of the spreading operator) of the GSC and the SC remains unchanged under these unitary transformations. 
Furthermore, the unitary invariance of the SC can also be used to provide a geometrical meaning of the SC when the Hamiltonian is an element of the $su(2), sl(2, R) $ or Heisenberg-Weyl Lie algebra.  However, at this point, it is to important to note that, we were able to provide such a geometrical interpretation only for the first order GSC, i.e., SC, and though for the class of Hamiltonians mentioned above, the GSC of order $m$ can be written as a polynomial function of the SC of degree $m$, we do not have any geometrical interpretation for these higher order complexities. 

Since here we are considering a unitary Hamiltonian time evolution, a natural question one can ask is: what is the amount by which different information-theoretic quantities associated with the initial state change due to this evolution?  To provide an example,  in \cite{Deffner}, an upper bound on the change of the Shannon entropy in a unitary evolution was obtained. Since, both the Shannon entropy and the SC depend on the probabilities of obtaining the time-evolved state in terms of the elements of a complete orthonormal basis (which in the first case can be any orthonormal basis and the Krylov basis in the latter case), it is also interesting to obtain such an upper bound on the change of the GSC. 
We obtain such a bound in section \ref{bounds}  in terms of the operator norm of the Hamiltonian and the function of the probability amplitudes in the Krylov basis, and discuss various physical significances of it. Specifically, we obtain an explicit expression for it in terms of the SC when the Krylov subspace is two-dimensional.  

In recent years, the concept of the Krylov operator and state complexity 
have gained a lot of attention due to their application in probing systems that show signatures of quantum chaos.  Specifically, as we have mentioned before, in the context of the complexity of the spread of a time-evolved state under unitary evolution,  the peak in the SC has been used in several works as a signature of level repulsion in the energy spectrum.  Keeping this fact in mind, we have shown in section \ref{contiumlimit} that the GSCs of all orders, in the limit, when the dimension of the Krylov subspace is large so that one can assume that the index $n$ is essentially a continuous variable $x$,  have a peak in their time evolution pattern before it reaches the corresponding saturation values. In fact, these peaks are more prominent in the plots for higher-order GSCs.  Apart from the assumption of the continuum limit, the main ingredients we have used to this end are the two-point correlation function of the GUE and a formula for the average of  Lanczos coefficients \cite{dumitriu2002matrix,Balasubramanian:2022dnj}.  Therefore, we believe that the peaks in the GSC are a generic signature of the chaotic nature of the energy spectrum of the Hamiltonian, and it will be interesting to verify this numerically.  However, one important point we should note in this context is that the average expression for the Lanczos coefficients distribution in eq. \eqref{LCrmt} is valid for a specific initial state $(1,0,0,\cdots 0)^T $, where these coefficients are written in terms of a basis in which the Hamiltonian generating the time evolution is a random matrix taken from the Gaussian ensemble under consideration. Though this approximation for the Lanczos coefficients is also argued to be a good approximation in the large $L$ limit for an initial maximally entangled infinite temperature thermofield double state (we have used this fact in section \ref{contiumlimit} to derive the expression for the GSC in the continuum limit for this initial state), it may not be a good approximation for other arbitrary initial states that are not related $(1,0,0,\cdots 0)^T $.  Therefore, it is important to remember that for such initial states, the GSC of the corresponding time-evolved state may not show a peak, even for a chaotic Hamiltonian.  

Before concluding this section, we briefly discuss a few possible interesting directions along which the results presented in the present manuscript can be extended. 
	
1. \textbf{Modification of the cost function with a higher order of probabilities.} Most of the works so far in the literature related to the spreading of a time-evolved state in the Krylov basis have used the cost function introduced in \cite{Balasubramanian:2022tpr}. However, as we have already seen towards the end of section  \ref{bounds},  one can define new and useful cost functions by modifying the original used to define the SC. E.g., one possible modification can be obtained by using a power of $k$, with $k=1,2,3, \cdots$ of the transition probabilities in the Krylov basis in the summation of eq. \eqref{cost}, where the expression for the  SC is restored when $k=1$. Of course, one needs to check whether these modified costs are also minimised in the Krylov basis at least up to a finite amount of time to justify their status as a measure of complexity associated with the wavefunction spreading. 
 
2. \textbf{Statistics of spreading in non-equilibrium quantum dynamics.} As we have mentioned previously in section \ref{statO}, when the initial state before the evolution is constructed from eigenstates of a given Hamiltonian and the time evolution is generated by the same Hamiltonian with different values of the parameters, the protocol can be thought of as a sudden quantum quench process. In section \ref{su2}, we have also discussed examples of models where the  Hamiltonians considered for analytical computations might arise. It will be interesting to study various properties of the statistics of spreading in some concrete many-body quantum systems in a sudden quench set-up numerically.  Specifically, one particular interest would be to consider chaotic Hamiltonians and see whether the peak in the SC  (as well as GCS),  which is usually marked as the signature of the chaotic nature of the energy spectrum, continues to be present when the initial state is constructed from the pre-quench Hamiltonian eigenstates.  We hope to report on this issue in detail in an upcoming work \cite{toappear}.  

Furthermore,  in principle, it is possible to think of a quench scenario where the spreading operator is the Hamiltonian before the quench. In that case, the initial state before the quench is the ground state of the pre-quench Hamiltonian, and the statistics in \eqref{SOsta} describes the statistics associated with the change in the energy of the pre-quench Hamiltonian. Since the pre-quench Hamiltonian would then essentially describe a harmonic oscillator of unit frequency, it is possible to evaluate the generating function  $G(\eta,t)$, and the GSCs analytically in a sudden quench protocol where one changes the frequency of the oscillator \cite{toappear}. 
  
We also point out that in this paper, we have considered time evolution generated by Hamiltonians that are time-independent, or from the perspective of quenches, these are sudden quenches. As the next step, one can consider complexity evolution and the statistics of spreading in the more 
general case of smooth quench protocols.  In such cases, one can control the rate of the quench and, therefore, study the scaling behaviour of various quantities with respect to this rate.
It would be interesting to study the complexity evolution in such a smooth quench protocol, and we hope to report on this in the future. 

3. \textbf{Evolution with initial mixed state.} In this paper we have considered the scenario where the initial state before the start of the evolution at $t=0$ is a pure state. One can consider a more general setup where the state before the evolution is a mixed state and study the statistics of the measurement of the spreading operator. In such cases, one possible approach would be to consider a purification of this density matrix by embedding it in a larger Hilbert space so that one can study the spread of the time-evolved wavefunction in this extended Hilbert space by using the usual formalism for the wavefunction spreading applicable for a pure state.  Specifically, when a sudden quench setup is considered, and the spreading operator is assumed to be the same as the initial pre-quench Hamiltonian, then the initial density matrix can be taken to be of the form, $\rho_{\text{ini}}=\sum_{n} p_n^0 \ket{K_n}\bra{K_n}$,  with $\sum_n p_n^0=1$, so that the pre-quench Hamiltonian commutes with the initial density matrix. In such cases, an initial measurement of $\mck$ before the quench gives a result $n$ with probability $p_n$ without destroying the coherence of the initial state, and therefore, one can easily define the statistics of the change of the spreading number under the quench by performing a second measurement of $\mck$ at an arbitrary time after the quench.


\begin{center}
	\bf{Acknowledgments}
\end{center}

We sincerely thank  Hugo A. Camargo and Viktor Jahnke for many insightful discussions, comments on a draft version of the manuscript,  and collaboration on related projects. We also thank Mohsen Alishahiha and Ankit Gill for very useful correspondence; Mitsuhiro Nishida and Bidyut Dey for comments on a draft version of this manuscript. The work of Kunal Pal is  supported by the YST Program at the APCTP through the Science and
Technology Promotion Fund and Lottery Fund of the Korean Government. This was also supported by the Korean Local Governments -
Gyeongsangbuk-do Province and Pohang City.
This work was supported by the Basic Science Research Program through the National Research Foundation of Korea (NRF) funded by the Ministry of Science, ICT \& Future
Planning (NRF2021R1A2C1006791) and the Al-based GIST Research Scientist Project
grant funded by the GIST in 2024. KYK was also supported by the Creation of the Quantum Information Science R\&D Ecosystem (Grant No.2022M3H3A106307411) through the
National Research Foundation of Korea (NRF) funded by the Korean government (Ministry of Science and ICT). Yichao Fu, Kunal Pal, and Kuntal Pal contributed equally to this paper and should be considered as co-first authors. 

\appendix
 \section{Long time average of generalised spread complexity}
In this appendix we briefly discuss the long-time averages of the moments of the characteristic function (eq. \eqref{CFSC})  or GSCs,  as well as cumulants of the spreading distribution,  since these provide important information regarding the late-time saturation value of the SC and  Krylov operator complexity  \cite{Rabinovici:2021qqt, Rabinovici:2022beu}. E.g., authors of ref.  \cite{Craps:2023ivc} recently used the long-time average of the SC to connect the idea of wavefunction spreading in the Krylov basis with the geometric notion of circuit complexity formulated by  Nielsen. 
	
The  long-time average of the GSC is defined as 
	\begin{equation}
		\bar{\mathcal{C}}_{m} =  \lim_{T \rightarrow \infty}  \frac{1}{T}\int_{0}^{T}  \mathcal{C}_{m} (t) ~dt~.
	\end{equation}
	Taking this limit in eq. \eqref{GSC2}, we get the time average value of the GSC to be 
	\begin{equation}\label{average1}
		\bar{\mathcal{C}}_{m} =  \sum_{a}  A_m (E_a)  ~ \rho_0 (E_a)~,
	\end{equation}
where $A_m (E_a) $ and  $\rho_0 (E_a)$ denote the diagonal elements of the generalised spreading operator and the initial state density matrix, respectively. Thus 
	\begin{equation}
		A_m (E_a)  =   \sum_n n^m |\braket{E_a | K_n}| ^2 =   \sum_n n^m \mathcal{P}_n^2 (E_a) ~ \rho_0 (E_a)~.
	\end{equation}
Therefore, the long-time average of the mean spreading, i.e. $\bar{\braket{n}}=\bar{\mathcal{C}} (t) $, and all its higher order generalisations (with higher values of $m \neq1$),
depend only on the diagonal component of the matrix $A_m$ (and the initial state density matrix).
	
Now using the expression for the GSC in eq. \eqref{GSC2}, the  variance in eq. \eqref{variance},  can be rewritten as 
	\begin{equation}
		\begin{split}
			(\Delta n (t))^2 = \mathcal{C}_2(t)- \mathcal{C}^2(t) \hspace{3 cm}\\
			= \sum_{ab} (A_2)_{ab} ~(\rho_0)_{ab} ~e^{i E_{ab}t}  
			- \sum_{abcd} (A_1)_{ab} ~ (A_1)^{*}_{cd} ~ (\rho_0)_{ab}~(\rho_0)^{*}_{cd} ~e^{i(E_{ab}+E_{cd})t}~.
		\end{split}
	\end{equation}
Here we have introduced the notations $ (A_m)_{ab} =  A_m(E_a, E_b)$, $(\rho_0)_{ab} = \rho_0 (E_a, E_b) $, and $E_{ab}= E_a-E_b$ for convenience. Assuming  the energy spectrum to be non-degenerate and the energy difference to be rational, after taking the long-time average of this expression  for the variance, we get 
	\begin{equation}
		\begin{split}
			\overline{(\Delta n)^2} =  \lim_{T \rightarrow \infty}  \frac{1}{T}\int_{0}^{T} (\Delta n (t))^2 dt \hspace{2cm}\\
			=\sum_{a} (A_2)_a (\rho_0)_a - \sum_{ab}  (A_1)_a (\rho_0)_a (A_1)_b (\rho_0)_b - \sum_{a\neq b}  |(A_1)_{ab} |^2 |(\rho_0)_{ab}|^2~.
		\end{split}
	\end{equation}
Therefore, in contrast with the long-time average of the mean spreading $\bar{\braket{n}}$ (see eq. \eqref{average1}), the long-time average of the variance of the spreading depends on both the diagonal and nondiagonal elements of the matrix version of $A_m$, written in the energy basis. 

\bibliography{reference}

\end{document}